\documentclass[twocolumn,showpacs,floatfix,
superscriptaddress,
longbiblography,
amsmath,amssymb,
aps,
prl]{revtex4-2}

\usepackage{graphicx}% Include figure files
\usepackage{dcolumn}% Align table columns on decimal point
\usepackage{amsmath}
\usepackage{bm}% bold math
\usepackage{appendix}
\usepackage{multirow}
\usepackage[bookmarks=true,colorlinks=true,urlcolor=blue,linkcolor=blue,citecolor=blue,breaklinks]{hyperref}
\usepackage[mathlines]{lineno}% Enable numbering of text and display math
\usepackage{inconsolata} % very nice fixed-width font included with texlive-full
\usepackage[usenames,dvipsnames]{color} % more flexible names for syntax highlighting colors
\usepackage{listings}

\usepackage{pdfpages} % include pdfs
\usepackage{pgffor} % for loops
\usepackage{xr} % referencing the supplement

\makeatletter
\AtBeginDocument{\let\LS@rot\@undefined}
\makeatother

\def\supplementfilename{supplement}

\pdfximage{\supplementfilename.pdf}
\def\numbersupplementpages{\the\pdflastximagepages}

\newif\ifarXiv
\arXivtrue 
\newcommand\underrel[2]{\mathrel{\mathop{#2}\limits_{#1}}}

\begin{document}

\preprint{APS/123-QED}

\title{Quantum logarithmic multifractality}% Force line breaks with \\

\author{Weitao Chen}
\affiliation{Department of Physics, National University of Singapore, Singapore.}
\affiliation{MajuLab, CNRS-UCA-SU-NUS-NTU International Joint Research Unit, Singapore.}
\affiliation{Centre for Quantum Technologies, National University of Singapore, Singapore.}
\author{Olivier Giraud}
\affiliation{MajuLab, CNRS-UCA-SU-NUS-NTU International Joint Research Unit, Singapore.}
\affiliation{Centre for Quantum Technologies, National University of Singapore, Singapore.}
\affiliation{Universit\'e Paris-Saclay, CNRS, LPTMS, 91405 Orsay, France.}
\author{Jiangbin Gong}%
\email{phygj@nus.edu.sg}
\affiliation{Department of Physics, National University of Singapore, Singapore.}
\affiliation{MajuLab, CNRS-UCA-SU-NUS-NTU International Joint Research Unit, Singapore.}
\affiliation{Centre for Quantum Technologies, National University of Singapore, Singapore.}
\author{Gabriel Lemari\'e}%
\email{lemarie@irsamc.ups-tlse.fr}
\affiliation{MajuLab, CNRS-UCA-SU-NUS-NTU International Joint Research Unit, Singapore.}
\affiliation{Centre for Quantum Technologies, National University of Singapore, Singapore.}
\affiliation{Laboratoire de Physique Théorique, Université de Toulouse, CNRS, UPS, France.}

\date{\today}% It is always \today, today,

\begin{abstract}
Through a combination of rigorous analytical derivations and extensive numerical simulations, this work reports an exotic multifractal behavior, dubbed ``logarithmic multifractality",  in effectively infinite-dimensional systems undergoing the Anderson transition.  In marked contrast to conventional multifractal critical properties observed at finite-dimensional Anderson transitions or scale-invariant second-order phase transitions,  in the presence of logarithmic multifractality, eigenstate statistics, spatial correlations, and wave packet dynamics can all exhibit scaling laws which are algebraic in the {\it logarithm} of system size or time. 
Our findings offer crucial insights into strong finite-size effects and slow dynamics in complex systems undergoing the Anderson transition,  such as the many-body localization transition.
\end{abstract}

\maketitle

%\tableofcontents

\emph{Introduction.\textemdash}
Determining the critical threshold of a continuous phase transition can be challenging. At finite system size, temperature, or time, such transition smoothens into a crossover, lacking singular behavior. While second-order phase transitions benefit from the scale-invariance property to discern their threshold \cite{wilson1983renormalization}, other transitions like Kosterlitz-Thouless (KT) type transitions \cite{kosterlitz2016kosterlitz} present additional complexities, including logarithmic finite-size effects \cite{hsieh2013finite}.

The Anderson localization transition, arising
from disorder and interference effects, is a well-studied
second-order phase transition in finite dimensions \cite{RevModPhys.80.1355, abrahams201050}. 
Researchers have employed various methods to determine the critical point \cite{castellani1986multifractal, PhysRevLett.102.106406,PhysRevLett.103.155703,PhysRevB.84.134209,PhysRevA.80.043626,PhysRevLett.105.090601,PhysRevLett.101.255702,PhysRevLett.87.056601,PhysRevB.95.094204, OhtsukiKawarabayashi,PhysRevA.94.033615,PhysRevLett.69.695,PhysRevB.82.161102,Kravtsov_2011,ALTSHULER2023169300,PhysRevA.95.041602,martinez2022coherent,PhysRevA.100.043612,PhysRevE.108.054127}. Central to this line of research, quantum multifractality is understood as one  key property depicting strong and scale-invariant spatial fluctuations of eigenstates at the critical point \cite{mandelbrot1974intermittent,mandelbrot1982fractal,falconer2004fractal}. The said quantum multifractality is investigated by moments $P_q \equiv \sum_i |\psi(i)|^{2q}$ of order $q$ of on-site eigenstate amplitudes $|\psi(i)|^{2}$ exhibiting an algebraic scaling behavior $P_q \sim N^{-D_q(q-1)}$ with the system size $N$. While $D_q=1$ for ergodic delocalization and $D_q=0$ for localization, $D_q$ for multifractal eigenstates is a non-trivial function of $q$ with $0<D_q<1$.

Recently, there has been considerable interest in the Anderson transition in infinite dimensionality (AT$^\infty$), e.g., on random graphs,  due to its analogy with the many-body localization (MBL) transition \cite{PhysRevB.34.6394,ZIRNBAUER1986375,PhysRevLett.67.2049,PhysRevLett.72.526,Monthus_2011,biroli2012difference,PhysRevLett.113.046806,PhysRevLett.117.156601,PhysRevB.94.220203,PhysRevB.94.184203,PhysRevB.96.214204,PhysRevLett.118.166801,PhysRevB.96.201114,PhysRevB.95.094204,KRAVTSOV2018148,PhysRevB.99.214202,PhysRevB.101.100201,Parisi_2020,PhysRevB.98.134205,PhysRevB.99.024202,PhysRevLett.125.250402,PhysRevB.105.094202,TIKHONOV2021168525,PhysRevResearch.2.012020,PhysRevB.105.174207,PhysRevB.106.214202,vanoni2023renormalization,baroni2023corrections,fyodorov1992novel,ADMirlin_1991}. There is an ongoing debate regarding the critical disorder value (even its existence) beyond which eigenstates become many-body localized \cite{PhysRevB.95.155129,PhysRevB.105.174205,PhysRevB.106.L020202,leonard2022signatures,PhysRevB.100.104204,PhysRevE.102.062144,PhysRevB.102.064207,PhysRevB.103.024203,PhysRevE.104.054105,PhysRevLett.127.230603,ABANIN2021168415,PhysRevLett.124.186601,Panda_2019,PhysRevB.102.100202}. Addressing this debate presents an outstanding question due to subtle and unavoidable finite-size effects and slow dynamics present in both the MBL transition and the AT$^\infty$ \cite{ABANIN2021168415,TIKHONOV2021168525}. A complete understanding of finite-size effects and slow dynamics necessitates a thorough study of quantum multifractality in complex systems.

One of the critical features of the AT$^\infty$ is its exotic multifractal properties, characterized by $D_q=0$ for $q>q^*$ and $D_q>0$ for $q<q^*$ \cite{RevModPhys.80.1355}, with a threshold $q^*=\frac12$ \cite{PhysRevLett.97.046803, gruzberg2011symmetries, gruzberg2013classification, PhysRevResearch.3.L022023}. 
However, the condition $D_q=0$ for $q>q^*$ does not reliably identify the transition point, as knowledge of \textit{how} $D_q$ vanishes with system size is essential.
For random regular and Erdős-Rényi graphs, it has been shown analytically that $P_2\sim (\ln N)^{-1/2}+P_2^{N=\infty}$, where $P_2^{N=\infty}>0$ signifies true localization behavior in the thermodynamic limit \cite{PhysRevLett.67.2049,PhysRevLett.72.526,PhysRevB.99.024202,fyodorov1992novel,ADMirlin_1991}. This type of critical behavior is termed ``critical localization''. By contrast, numerical simulations have suggested another possibility where $P_2$ is algebraic in ln N, also compatible with $D_2=0$ \cite{PhysRevLett.118.166801,PhysRevResearch.2.012020,PhysRevB.106.214202, chen2023anderson}. This behavior hints at the possibility of $P_q \sim {(\ln N)}^{f(q)}$, namely, a power function of $\ln N$ with a $q$-dependent exponent $f(q)$. This second scaling behavior is dubbed below ``logarithmic (or log-) multifractality''.

Distinguishing between the above-mentioned  two critical behaviors is crucial. Critical localization resembles a Kosterlitz-Thouless (KT) behavior in terms of $P_2$: Throughout the localized phase, $P_2^{N=\infty}$ remains finite until a sudden drop to zero in the delocalized phase, accompanied by characteristic logarithmic finite-size effects at the transition \cite{PhysRevB.99.024202}. Markedly different, log-multifractality entails the following scenario: localized wave functions on tree-like graphs explore only a few rare branches \cite{PhysRevLett.118.166801,PhysRevResearch.2.012020,PhysRevB.106.214202}. This support spans $\ln N$ sites. At the transition, wavefunctions become multifractal on this support. This represents a log-scale invariance not yet explored, to the best of our knowledge.  

Based on both analytical and numerical simulations, the present work demonstrates the existence of log-multifractality.   Our starting points are the Power-law Random Banded Matrix (PRBM) ensemble \cite{PhysRevE.54.3221, PhysRevB.62.7920,PhysRevE.98.042116} and the Ruijsenaars-Schneider (RS) ensemble \cite{PhysRevLett.103.054103,PhysRevE.84.036212,PhysRevE.85.046208}, previously introduced to explore quantum multifractality thanks to their amenability to analytical treatment \cite{PhysRevE.54.3221, PhysRevB.62.7920, L.S.Levitov_1989,  PhysRevLett.64.547, Monthus_2010, PhysRevLett.106.044101, Kravtsov_2011,Yevtushenko_2007,PhysRevB.82.161102, PhysRevE.98.042116, PhysRevLett.103.054103, PhysRevE.84.036212, PhysRevE.85.046208}. %\og{(GL: add a sentence saying that the bandwidth $b$ controls the ``dimensionality'' going from weak ($b \gg 1$) to strong multifractality ($b\ll 1$), but that this is not true strong multifractality, as $D_2 \propto b$ is always finite, and thus conventional multifractality.)} 
Existing efforts have focused on developing various random matrix ensembles to capture some aspects of AT$^\infty$. Investigations of variants of the Rosenzweig-Porter model have revealed multifractal phases and slow dynamics \cite{Kravtsov_2015, 10.21468/SciPostPhys.6.1.014, PhysRevResearch.2.043346, von2019non, Truong_2016, PhysRevE.98.032139, Monthus_2017, Amini_2017, PhysRevB.103.104205, kravtsov2020localization, PhysRevResearch.2.043346, 10.21468/SciPostPhys.11.2.045}. However, so far these models fall short of capturing the spatial properties of eigenstates crucial to describe delocalization-localization transition \cite{10.21468/SciPostPhys.11.2.045}. %, mainly due to the assumption of uniform variance for all off-diagonal matrix elements.
For our purpose here we introduce a variant of the PRBM ensemble \cite{PhysRevE.54.3221, PhysRevB.62.7920} by incorporating a specific decay of the off-diagonal matrix elements. 
To facilitate our extensive numerical simulations, we also introduce a unitary model akin to the so-called kicked rotor and RS models \cite{PhysRevLett.94.244102, PhysRevE.84.036212,PhysRevE.85.046208}. Also amenable to analytical treatment, the unitary model introduced here allows us to reach very large system sizes and therefore to investigate dynamical evolution in the regime required to validate the predicted log-multifractal properties. After validating log-multifractality through the algebraic behavior of $P_2$ in $\ln N$ both analytically and numerically, we explore the slow decay of eigenstate spatial correlations and derive characteristics of wave packet dynamics. As one remarkable consequence of log-multifractality, we find that the return probability exhibits an algebraic decay with $\ln t$ rather than with the time variable $t$ itself.
%Lastly, we extend our random matrix model to describe the %"critical localization" scenario for strong multifractality.

\begin{figure*}
%\centering
\includegraphics[width=0.32\textwidth]{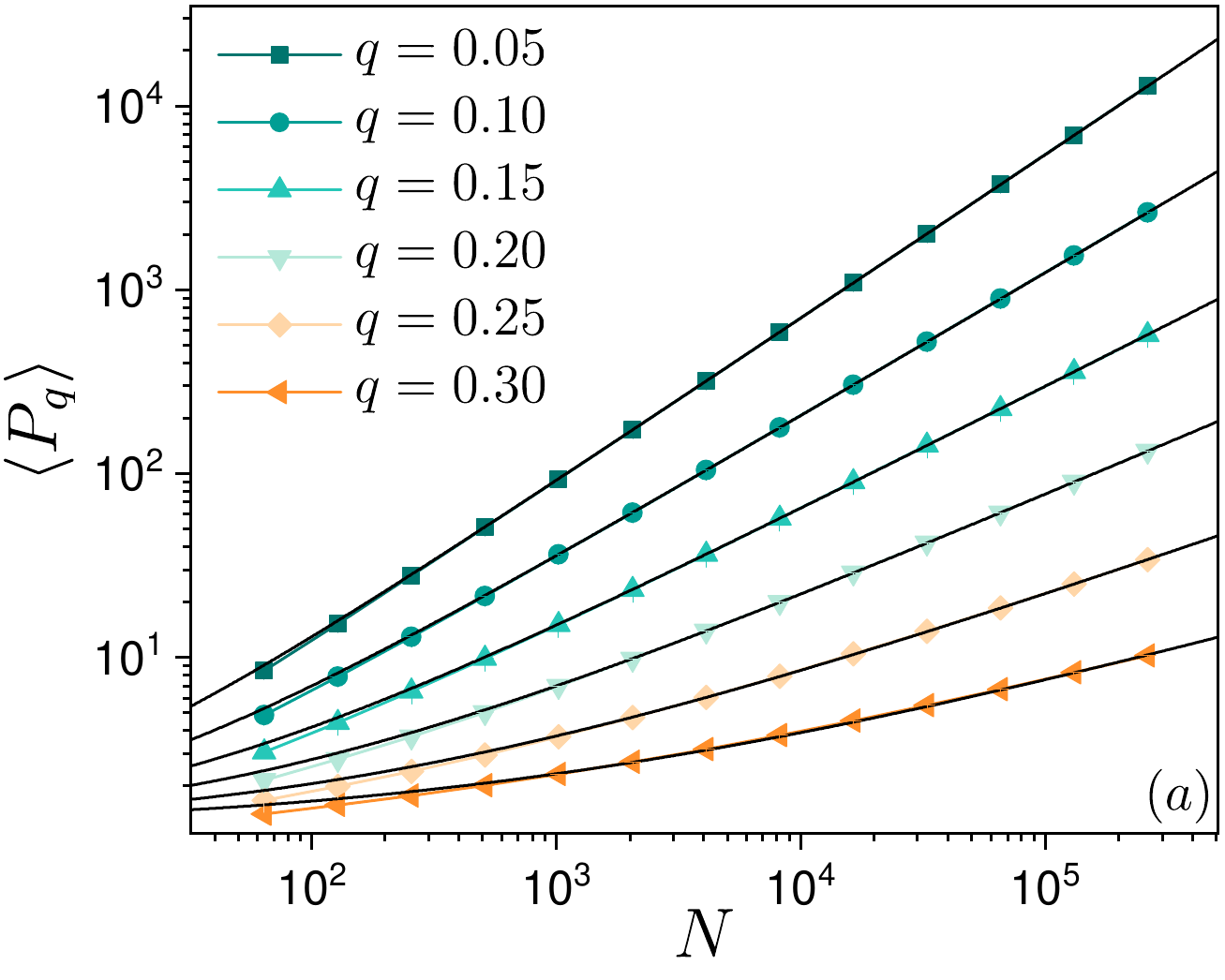}
\includegraphics[width=0.32\textwidth]{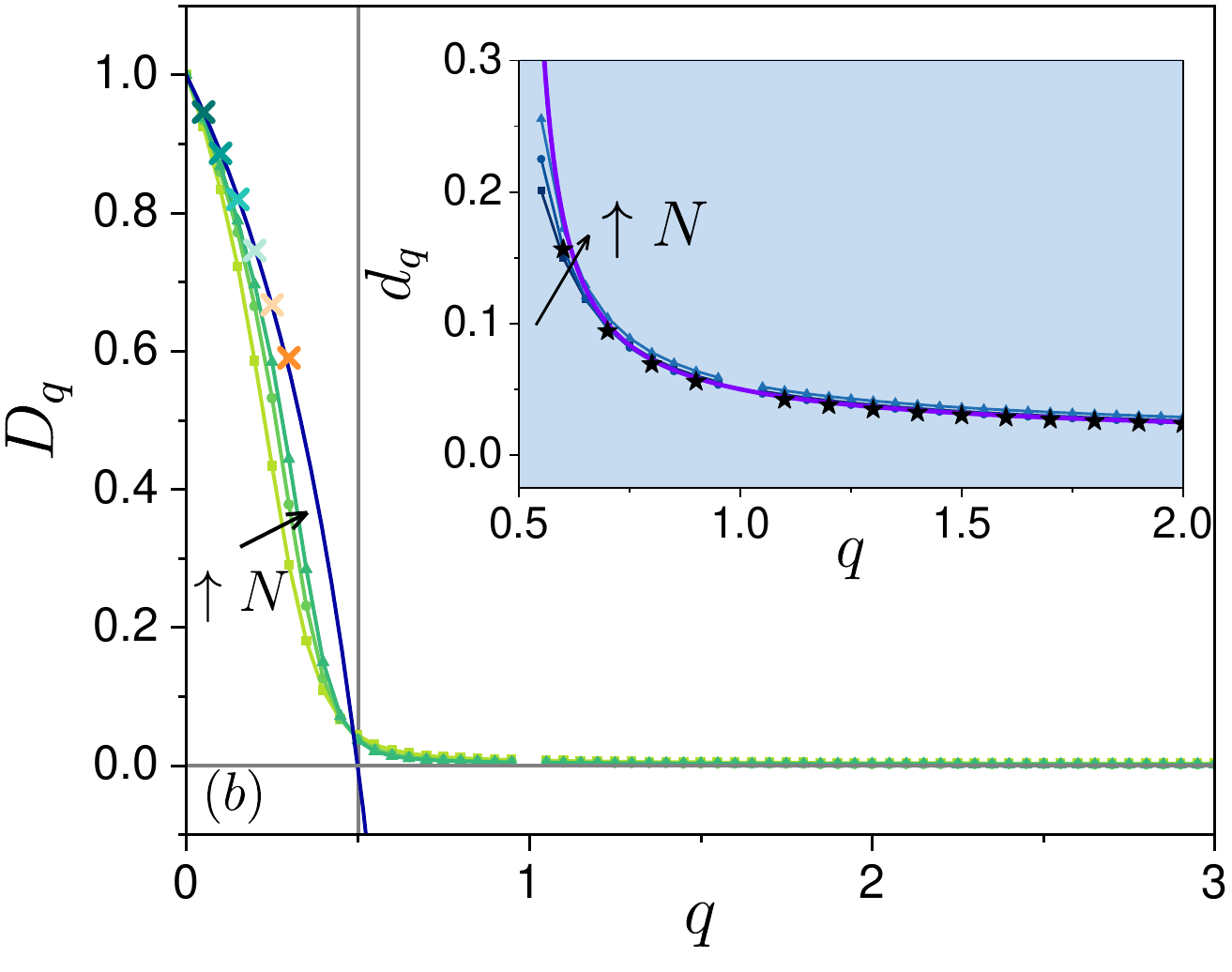}
\includegraphics[width=0.33\textwidth]{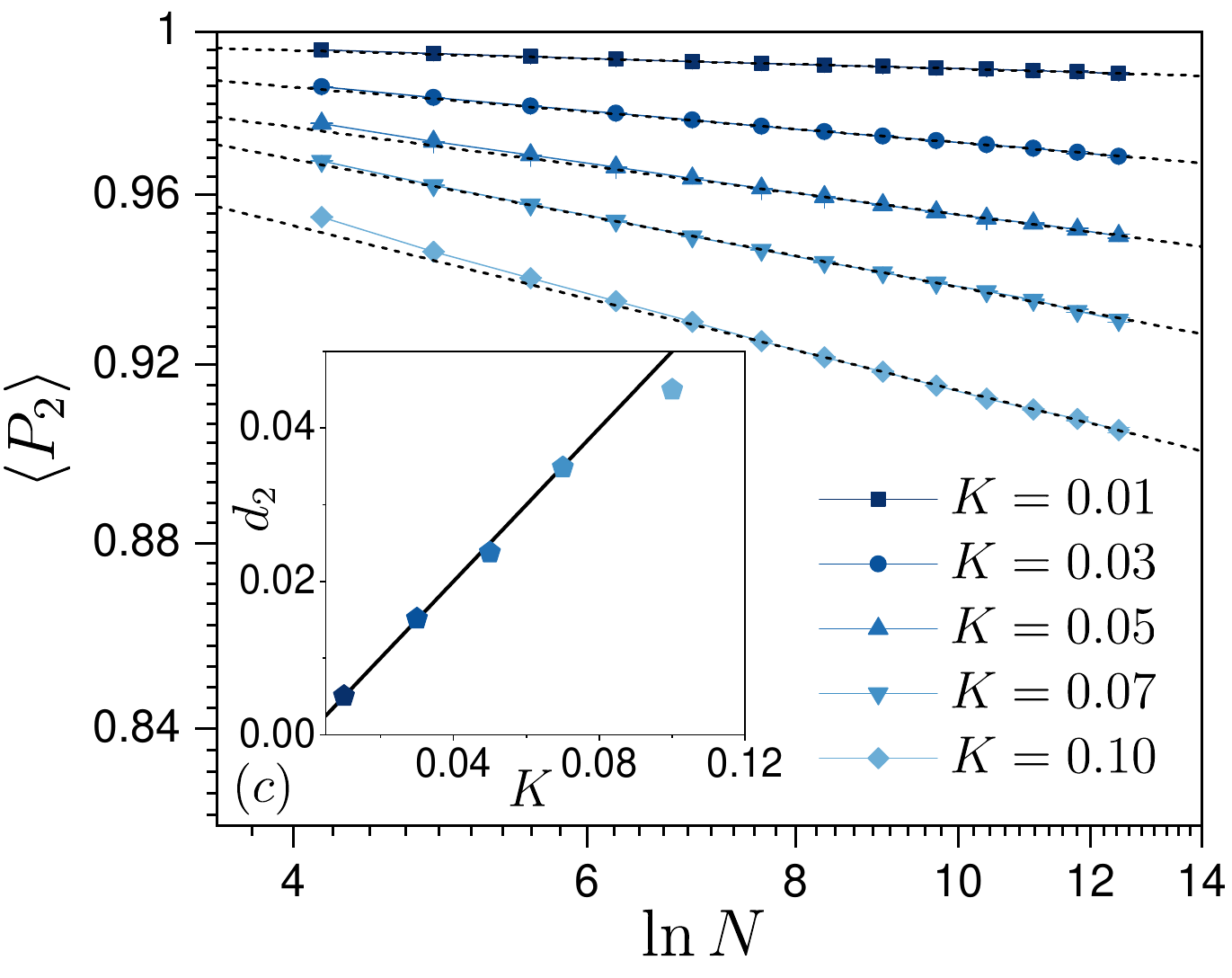}

    \caption{\label{fig1}  Eigenstate multifractality in the SRUM model. (a) Conventional multifractality for moments $P_q$ with $q<\frac12$ as predicted by Eq.~\eqref{eq:P2SMKR_} for $K=0.05$ and different $q$ values as indicated by the labels. The black dashed lines are fits by Eq.~\eqref{eq:P2SMKR_} with $A_{q}$ and $D_q$ two fitting parameters. (b) Multifractal dimension $D_q$ vs $q$ for $K=0.05$. The finite-size estimate $D_q\equiv [\log_2 \langle P_q(N/2)\rangle-\log_2 \langle P_q(N) \rangle]/[q-1]$, represented by green symbols (lines are an eyeguide) for system sizes $N=2^{10},2^{14},2^{18}$, converges slowly to the theoretical prediction $D_q=(2q-1)/(q-1)$ for $q<1/2$. The crosses indicate the $D_q$ values obtained from the fits from Eq.~\eqref{eq:P2SMKR_} represented in panel (a), which incorporate the log-corrections and agree perfectly well with $D_q=(2q-1)/(q-1)$.
    Inset: Log-multifractal dimension $d_q$ (computed as $[\ln \langle P_q(N/2)\rangle-\ln \langle P_q(N) \rangle]/[(q-1)(\ln \ln N-\ln \ln \frac{N}{2})]$) as a function of $q$ for system sizes $N=2^{10},2^{14},2^{18}$. $d_q$ converges at large $N$ to the non-trivial analytical law \eqref{eq:P2SMKR} (violet line). Star symbols are fitted $d_q$ values shown in panel (c). (c) Log-multifractality for moments with $q>\frac12$, well described by Eq.~\eqref{eq:P2SMKR}. Different curves correspond to different $K$ values as indicated by the labels. The black dashed lines are power-law fits $\langle P_2\rangle = c (\ln N)^{-d_2}$ with $c$ and $d_2$ two fitting parameters. Inset: Comparison of the log-multifractal dimension $d_2$ obtained from fitting (symbols) and the analytical prediction Eq.~\eqref{eq:P2SMKR} (black solid line). Disorder averaging ranges from $360,000$ realizations for $N=2^6$ to $1,800$ realizations for $N=2^{18}$. Error bars are smaller than symbol size.  } 
\end{figure*}

\emph{Models.\textemdash} We start with the construction of a variant of the PRBM ensemble, namely, the  Strongly-multifractal  Random Banded Matrix (SRBM) ensemble, defined as an ensemble of $N\times N$ Hermitian matrices $\hat{H}$ whose entries $H_{ij}$ are independent Gaussian random variables with  mean $\langle H_{ij}\rangle=0$ and variance $\langle|H_{ij}|^2\rangle = \beta^{-1}$ for $i=j$ and  
\begin{equation}
\label{defSRBM}
\langle|H_{ij}|^2\rangle =\frac{1}{1+[|i-j|\ln(1+|i-j|)/b]^2}
\end{equation}
for $i\neq j$.
Here $\beta$ is the Dyson index for the orthogonal ($\beta=1$) and unitary ($\beta=2$) classes, and $b>0$ is a real parameter. To mitigate boundary effects in numerical simulations, we replace the term $|i-j|$ with $\sin(\pi|i-j|/N)/(\pi/N)$. In the limit $|i-j|\gg b$, the amplitude (standard deviation) of the off-diagonal elements decays as
\begin{equation}\label{eq3}
   \sqrt{ \langle|H_{ij}|^2\rangle}\simeq b \, (|i-j|\ln|i-j|)^{-1} \;.
\end{equation}
The decay behavior in Eq.~\eqref{eq3} represents the limiting case $a\to 1, a>1$, of the PRBM ensemble, where $\sqrt{\langle|H_{ij}|^2\rangle} \simeq (b/|i-j|)^a$ and the critical value $a=1$ distinguishes between a delocalized phase ($a<1$) and a localized phase ($a>1$) (more precisely a non-critical strongly multifractal phase with $q^*<\frac12$) \cite{PhysRevResearch.3.L022023}. Indeed we have $\lim_{\epsilon\rightarrow0}|i-j|^{1+\epsilon}\simeq |i-j|(1+\epsilon \ln|i-j|)$ for $\epsilon \ll 1$, that is,  the decay behavior \eqref{eq3} of the far-off-diagonal elements of SRBM can be roughly understood as the result of keeping the first-order term in $\ln|i-j|$ when taking the $a=1+0^{+}$ limit of the PRBM.  The long-range decay described by Eq.~\eqref{eq3} with logarithmic dependence on $|i-j|$ will be seen to play a pivotal role in inducing log-multifractality.

In addition to the above random Hermitian matrix ensemble, we also consider a unitary ensemble, named Strongly-multifractal Random Unitary Matrix (SRUM) ensemble.  The SRUM ensemble can be seen as a variant of the so-called kicked rotor model in quantum chaos and the RS models \cite{CHIRIKOV1979263,IZRAILEV1990299,PhysRevLett.94.244102,PhysRevE.84.036212,PhysRevE.85.046208}. %\og{(GL: cite also Wang and Garcia-Garcia, RS papers, etc.)}. 
The SRUM ensemble is comprised of random unitary matrices
\begin{equation}
\begin{split}
      U_{ij}=e^{i\Phi_i}\sum_{k=1}^{N}F_{ik}e^{-iKV(2\pi k/N)}F_{kj}^{-1},
\end{split}
\end{equation}
where $V(x) =   \ln\left[-1/\ln(\lambda|\sin\frac{x}{2}|)\right]$ for $x\in [0,2\pi)$, $V(x+2\pi)=V(x)$ and the Fourier transform $F_{jk}=e^{2i\pi jk/N}/\sqrt{N}$. The parameter $\lambda$ is set to $\lambda=0.9$ to avoid the singularity of $V(x)$ at $x=\pi$. $\Phi_{i}$ are random phases uniformly distributed over $[0,2\pi)$.  Due to the singular behavior of $V(x)$ when $x\rightarrow 0 \; (2\pi)$, the amplitudes of the matrix elements of $U_{ij}$ decay as $|U_{ij}|\simeq K/(2r\ln r)$ for large $r\equiv|i-j|$; this is the same behavior as $\sqrt{ \langle|H_{ij}|^2\rangle}$ in Eq.~\eqref{eq3} with $b$ replaced by $K/2$. This is a further justification of why we have used a singular $V(x)$ as above.

\emph{Log-multifractality\textemdash} To examine how log-multifractality might emerge in systems with long-range coupling,  we analytically compute $\langle P_q\rangle$ by treating the off-diagonal matrix elements in SRBM and SRUM models as perturbation \cite{PhysRevE.84.036212,PhysRevE.85.046208}.  In the case of SRUM, the unperturbed matrix is diagonal and at order zero eigenstates are $|\psi_{i}^{0}\rangle=|i\rangle$. At first order in $K$, eigenfunctions are given by $ |\psi_{i}^{1}\rangle=|i\rangle+\sum_{i\neq j} |j\rangle U_{ji}/(e^{i\Phi_i}-e^{i\Phi_j})$. The moments $P_q$ then read
\begin{equation}\small\label{eq9}
\begin{split}
      \langle P_q\rangle&\simeq 1+\sum_{i\neq j}\langle|e^{i\Phi_i}-e^{i\Phi_j}|^{-2q}\rangle|U_{ij}|^{2q}\,.
\end{split}
\end{equation}
%where $e^{i\Phi_i}$ and $e^{i\Phi_j}$, the unperturbed eigenvalues, are the diagonal elements $U_{ii}$. 
Disorder averaging
%, denoted $\langle \rangle$, 
can be performed as $\langle|e^{i\Phi_i}-e^{i\Phi_j}|^{-2q}\rangle=(2\pi)^{-1}\int_0^{2\pi}|1-e^{i\phi}|^{-2q}d\phi$. For $q<\frac12$, this integral converges and Eq.~\eqref{eq9}  yields
%\begin{equation}\small\label{eq:P2SMKR_}
%\begin{split}
%      \langle P_{q}\rangle& \simeq1+ \left(\frac{K}{4}\right)^{2q}\frac{2\Gamma(\frac{1}{2}-q)}{\sqrt{\pi}\Gamma(1-q)}\frac{N^{1-2q}(\ln N)^{-2q}}{1-2q}.
%\end{split}
%\end{equation}
\begin{equation}\label{eq:P2SMKR_}
      \langle P_{q}\rangle \simeq1+K^{2q} A_{q} (\ln N)^{-2q} N^{-D_q(q-1)},\quad D_q=\frac{2q-1}{q-1}
\end{equation}
with $A_{q}$ a constant. For large $N$, we approach the conventional multifractal behavior $P_q\sim N^{-D_q(q-1)}$. 
%However, the log-correction $(\ln N)^{-2q}$ implies a log-slow convergence of $q^{*} \simeq 1/[2(1+(\ln N)^{-1})]$ towards its critical value $q^*=\frac12$ when $N \rightarrow \infty$. This finding supports the notion of $q^*$ having a KT flow in the AT$^\infty$, see \cite{PhysRevB.106.214202}.

Importantly, the above disorder averaging diverges when $q>\frac12$, necessitating more advanced treatments. In this context, we employ Levitov renormalization \cite{L.S.Levitov_1989,PhysRevLett.64.547}, known for its effectiveness in the PRBM and RS ensembles \cite{PhysRevB.62.7920, PhysRevE.84.036212}. This approach considers contributions from all $2\times2$ sub-matrices of $U$. %Suppose every $2\times2$ sub-matrices $U_{mn}$ gives one pair of eigenvectors $(u_\sigma^{mn},v_\sigma^{mn})$ with $\sigma=\pm$ denotes different eigenvectors, the contribution of these eigenvectors to $\langle P_q\rangle$ can be computed as $N^{-1}\sum_{m<n}\sum_{\sigma=\pm}\langle|u_\sigma|^{2q}+|v_\sigma|^{2q}-1\rangle$ where the $-1$ term is introduced to eliminate the zero-order contribution.
By calculating the first-order contribution in this manner, disorder averaging converges, yielding the following expression for $q>\frac12$:
%. Calculating the eigenpairs of $U_{mn}$ explicitly and summing over the index $m$ and $n$, the prediction of $\langle P_q\rangle$ for $q>\frac12$ is achieved as
%\begin{equation}\label{eq:P2SMKR}
%\begin{split}
%      \langle P_{q}\rangle\simeq  (\ln N)^{-\frac{K\Gamma(q-\frac{1}{2})}{\sqrt{\pi}\Gamma(q-1)}}.
%\end{split}
%\end{equation}
\begin{equation}\label{eq:P2SMKR}
\begin{split}
      \langle P_{q}\rangle\sim  (\ln N)^{-d_q (q-1)}\;, \quad d_q = \frac{K\Gamma(q-\frac{1}{2})}{\sqrt{\pi}\Gamma(q-1) (q-1)}. \; 
\end{split}
\end{equation}
Similar treatments apply to SRBM \cite{LongPaper}. %Notably, an alternative technique exists for handling divergence when $q>\frac12$, beyond the Levitov renormalization, using the renormalization approach based on the generating function of the weighted L\'evy sums \cite{Monthus_2010}. For detailed calculations of this approach, refer to Ref. \cite{LongPaper}. The final analytical results 
For the orthogonal class ($\beta=1$),  we find that $\langle P_q \rangle$ is given by Eq.~\eqref{eq:P2SMKR_} for $q<\frac12$ and Eq.~\eqref{eq:P2SMKR} for $q>\frac12$, with $K$ replaced by $4b$.  In this regard, results from SRUM and SRBM fully echo with each other. 
Equation (\ref{eq:P2SMKR}) shows that our models display log-multifractality, as $P_q$ algebraically scales with $\ln N$ rather than $N$, and provides an explicit expression for the corresponding multifractal exponent.

%as follows:
%\begin{equation}\label{eq:P2SRBM}
%\begin{split}
%     \langle P_{q}\rangle&\simeq(\ln N)^{-\frac{\Gamma(q-\frac12)}{\Gamma(q-1)}\frac{4b }{\sqrt{\pi}}}
%\end{split}
%\end{equation}
%for $q>\frac12$ and
%\begin{equation}\small\label{eq:P2SRBM_}
%\begin{split}
%     \langle P_{q}\rangle&\simeq1+ \frac{2b^{2q}}{\pi(1-2q)}\Gamma(\frac12+q)\Gamma(\frac12-q) N^{1-2q}(\ln N)^{-2q}
%\end{split}
%\end{equation}
%for $q<\frac12$. 

%Our analytical predictions reveal two distinct signatures of strong multifractality that extend beyond conventional multifractality: i) For $q>\frac12$, moments scale as $\langle P_q\rangle\sim (\ln N)^{-d_q(q-1)}$, with nontrivial {\it strongly} multifractal dimensions $d_q$, while the {\it conventional} multifractal dimensions $D_q$ vanish, as described by Eq.~(\ref{eq:P2SMKR}) and Eq.~(\ref{eq:P2SRBM}). ii) For $q<\frac12$, there is a volumic scaling of $\langle P_q\rangle$ with system size $N$, incorporating corrections in $\ln N$, as described by Eq.~(\ref{eq:P2SMKR_}) and Eq.~(\ref{eq:P2SRBM_}); as $\ln N \rightarrow\infty$, this scaling goes to the conventional multifractal behavior with multifractal dimensions 
%\begin{equation}
%\label{dqtrivial}
%D_q=\frac{2q-1}{q-1},\qquad q<\frac12. 
%\end{equation}

To validate the analytical predictions \eqref{eq:P2SMKR_}--\eqref{eq:P2SMKR}, especially the algebraic behavior in $\ln N$, reaching large system sizes is essential. This is much easier to achieve in the SRUM case.
%Because SRBM is modeled by a full matrix, the SRUM case is much more convenient in numerical simulations. 
Indeed, implementing a sparse diagonalization approach assisted by a polynomial filter \cite{luitz2021polynomial}, we are able to  explore system sizes as large as $N=2^{18}$ with high number of random realizations. In Fig.~\ref{fig1}, the left panel illustrates the conventional multifractal behavior of moments $P_q$ with $q<\frac12$, fitting well with Eq.~\eqref{eq:P2SMKR_}. The multifractal dimension $D_q$ is displayed as a function of $q$ in the middle panel, vanishing for $q>\frac12$. The right panel showcases the log-multifractality of moments $P_q$ with $q>\frac12$, fitting effectively with Eq.~\eqref{eq:P2SMKR}. The log-multifractal dimension $d_q$ exhibits a non-trivial dependency on $q$ and $K$, well-accounted for by Eq.~\eqref{eq:P2SMKR}. These results are also verified for SRBM case \cite{LongPaper}, albeit at a higher computational cost.

\begin{figure*}
\includegraphics[width=0.39\textwidth]{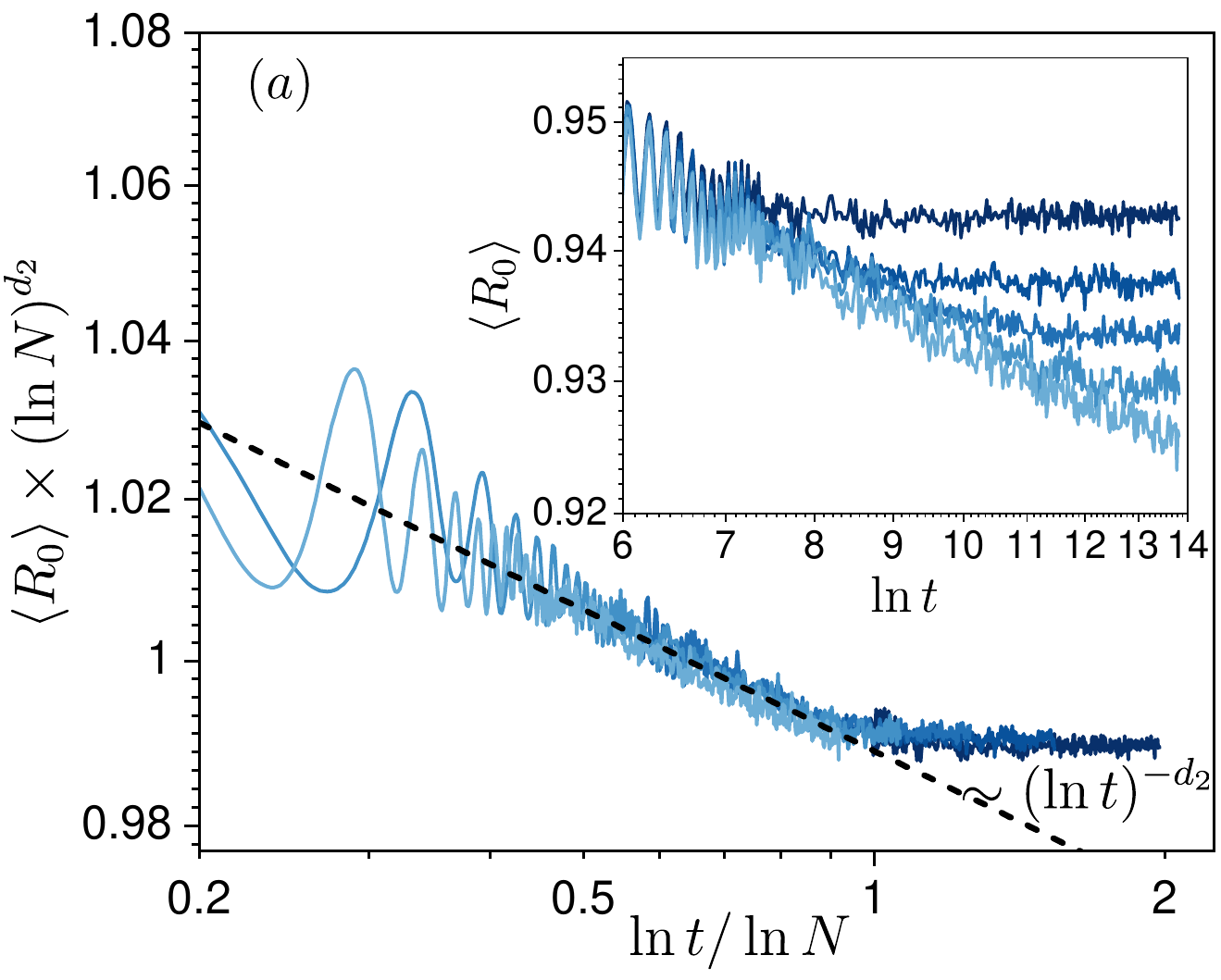}% 
\hfill
\includegraphics[width=0.56\textwidth]{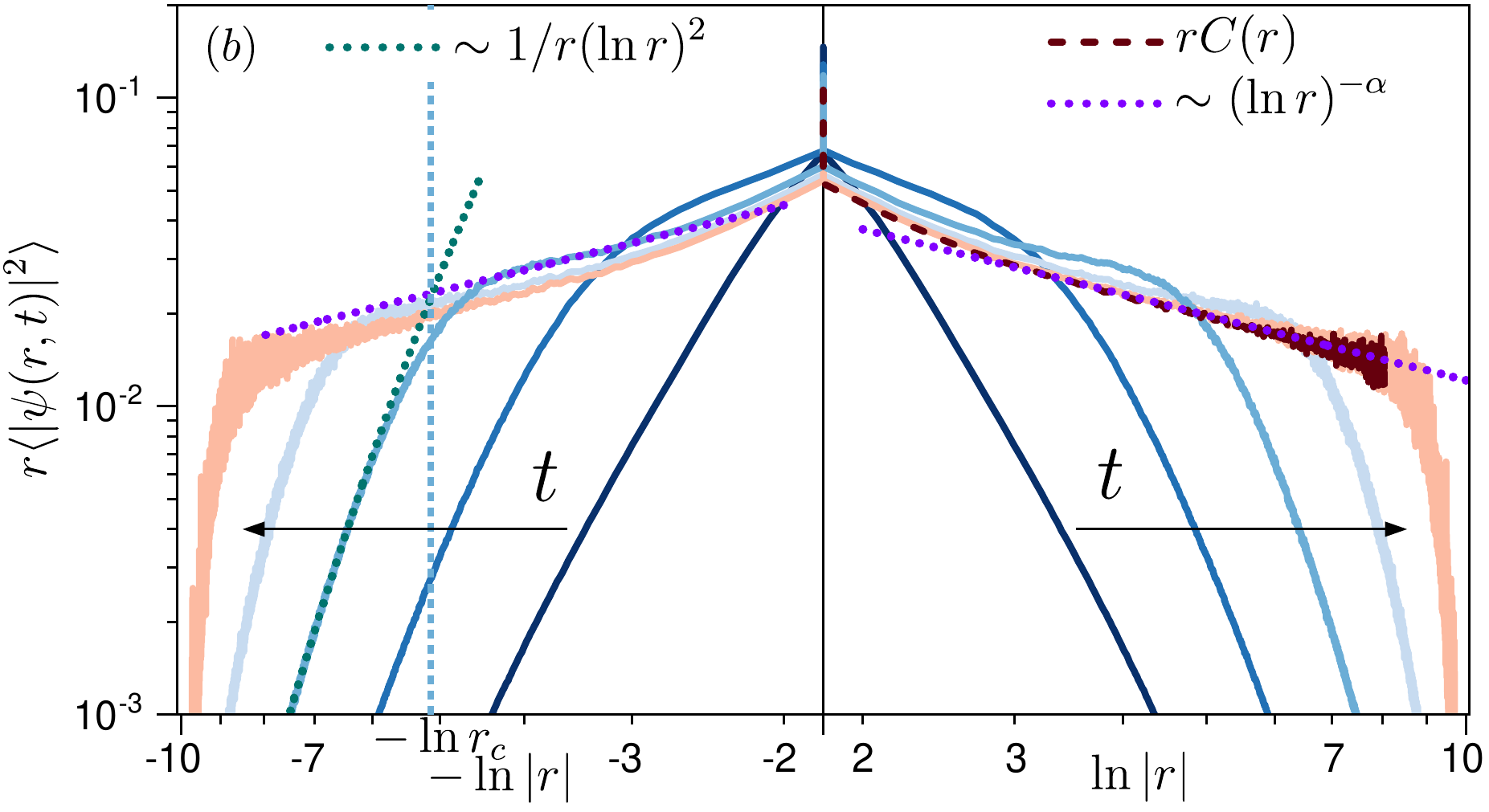}% 
\caption{\label{FigDynamics}
Slow wave packet dynamics for the SRUM with initial state $\psi(r, t=0) = \delta_{r,0}$, in connection with log-multifractality. (a) Illustration of the scaling property in Eq.~\eqref{eq:scaR0} for the return probability $\langle R_0 \rangle$ at $K=0.05$. The data from various system sizes and times $t \in [10, 10^6]$ collapse onto a single scaling curve when $\langle R_0 \rangle \times (\ln N)^{d_2}$ is plotted against the scaled time $\ln t/\ln N$. The black dashed line represents a power-law fit, $\langle R_0 \rangle \sim (\ln t)^{-d_2}$, giving $d_2\approx0.025$ in perfect agreement with the analytical prediction $d_2=K/2$ and the value extracted from $P_2$. Inset: corresponding raw data for $\langle R_0 \rangle$ with, from top to bottom, $N=2^7,2^9,\dots,2^{15}$. Results are averaged over a range from $18,000$ realizations for $N=2^7$ to $7,200$ realizations for $N=2^{15}$. (b) Average probability distribution of the wave packet at different times plotted as $r\langle|\psi(r,t)|^2\rangle$ vs. $\ln r$ in a symmetrical log-log scale, with $K=1.0$. Curves from dark blue to pale orange correspond to evolution times $t=10,77,1668,35398,10^5$. The violet dotted line fits the behavior expected at short distances $r \ll r_c$, corresponding to the correlation function $C(r)$ (shown by the maroon dashed line), giving $\alpha\approx0.7$; while the green dotted line fits the one for $r\gg r_c$, see Eq.~\eqref{eqpsi2}. The vertical dashed line locates the crossover $r_c$. Results are averaged over $720,000$ ($36,000$ for $C(r)$) disorder configurations with system size $N=2^{15}$ ($N=2^{16}$ for $C(r)$). }
\end{figure*}

%Another important probe of multifractality is the average correlation function of the eigenstates, which we find behaves as
%\begin{equation}\label{eq:cor}
%C(r) \equiv \langle \sum_{i=1}^{N} |\psi_j(i)|^2 |\psi_j(i+r)|^2 \rangle \sim\frac{1}{r(\ln r)^{\alpha}}
%\end{equation}
%with $0<\alpha<1$. This behavior is illustrated in Fig.~\ref{FigDynamics}(b) and can be understood as follows. For the conventional multifractality, $C(r) \sim r^{D_2-1}$ \cite{RevModPhys.80.1355}. In our case, this formula should be amended in two ways: first $r$ is replaced by $\ln r$ as we have multifractality in $\ln N$, see Eq.~\eqref{eq:P2SMKR}. Second, the support of the wavefunction is nontrivial, contrary to e.g.~the PRBM case. Indeed, only a fraction $1/r$ of sites $i$ participates to the sum in Eq.~\eqref{eq:cor}. This is analogous to the delocalization along a single branch over the exponentially many, which arises at the AT$^\infty$ \cite{biroli2012difference, PhysRevLett.118.166801, PhysRevResearch.2.012020, PhysRevB.106.214202}. This nontrivial support can be taken into account by including the factor $1/r$ to $C(r)$ \cite{PhysRevA.35.4907,PhysRevLett.66.247}, leading to: $C(r)\sim 1/\left[r(\ln r)^{1-d_2}\right]$. This suggests $\alpha = 1 - d_2$, a relation we have observed in the SRBM model but found deviations to in the SMKR model. This can be corrobarated by considering more general correlation functions, see \cite{LongPaper}.

We now turn to the average correlation function of eigenstates, $C(r) \equiv \langle \sum_{i=1}^{N} |\psi(i)|^2 |\psi(i+r)|^2 \rangle$, a key multifractality probe. It exhibits an exotic behavior given by $C(r) \sim (1/r) \times (\ln r)^{-\alpha}$, with $0<\alpha<1$. This behavior, depicted in Fig.~\ref{FigDynamics}(b) (red curve) is a distinctive consequence of log-multifractality when compared with  $C(r) \sim r^{D_2-1}$ found in conventional multifractality \cite{RevModPhys.80.1355}. This is further illustrated in Figs.~S1 and S2 of the Supplementary Material \cite{sup}. To roughly explain how these two different correlation functions differ from each other,  some additional results suggest that the fractional power function of $r^{D_2-1}$ in conventional multifractal cases is replaced by a power function of $\ln r$ to reflect multifractality in $\ln N$ (Eq.~\eqref{eq:P2SMKR}).  The additional $1/r$ factor in the $C(r)$ obtained here further indicates nontrivial wavefunction support.  More analysis on this side finding will be presented elsewhere~\cite{LongPaper}.
%further introduces a $1/r$ factor \cite{PhysRevA.35.4907,PhysRevLett.66.247}. %Incorporating these, $C(r)\sim 1/\left[r(\ln r)^{1-d_2}\right]$, suggesting $\alpha = %1 - d_2$. While SRBM aligns with this, we observe deviations in SMKR with $\alpha = 1 - d_2$%indicate the need for considering more general correlation functions, detailed in 
 %As a rough qulitative understanding of the exotic eigenstate spatial correlation %associated with log-multifractality,

%\og{Put part of it in the supmat} For the SRBM model, we observe that $\alpha=1-d_2$, i.e.,  $C(r)\sim 1/\left[r(\ln r)^{1-d_2}\right]$, indicating that $C(r)$ follows the slower decay compared to the typical decay of matrix elements $\sqrt{ \langle|H_{ij}|^2\rangle} \sim 1/\left(|i-j|\ln|i-j|\right)$. This slower decay is a consequence of the enhanced correlation by the fluctuations induced by strong multifractality, and it can be seen as an infinite-dimensional extension of the correlation function in the PRBM model, where correlations decay as $C(r) \sim r^{D_2-1}$ \cite{PhysRevA.35.4907}.

\emph{Wave packet dynamics.\textemdash} 
The dynamics of a wave packet initialized at a single site $\psi(r, t=0) = \delta_{r,0}$ also encodes rich information on quantum multifractality \cite{PhysRevE.108.054127,PhysRevE.86.021136,PhysRevB.82.161102,PhysRevA.95.041602, PhysRevA.100.043612}. 
In the context of conventional multifractality, the return probability $R_0(t)   \equiv|\langle \psi(t)|\psi(0)\rangle|^2$ exhibits a power-law decay with time $t$, $\langle R_0 \rangle\sim t^{-D_2}$, with an exponent given by the multifractal dimension $D_2$ \cite{PhysRevLett.69.695,PhysRevB.82.161102, Kravtsov_2011}. To analyze $R_0(t)$ in the case of log-multifracality, we
%Here, with log-multifracality shown above, it becomes intriguing to investigate how the return probability $\langle R_0 \rangle$ encodes distinctive features of log-multifractality. In our analytical treatment, we 
focus on the orthogonal SRBM model with Dyson index $\beta=1$.   We adapt the analytic expression for the return probability that was obtained for the PRBM case in the limit $b\ll 1$ by means of a supersymmetric virial expansion \cite{Kravtsov_2011}
 $\langle R_0\rangle=\langle R_0^{(0)}\rangle+\langle R_0^{(1)}\rangle+\dots$ in terms of successive orders of the parameter $b$, with $\langle R_0^{(0)}\rangle=1$. Adapting Eq.~(B.1) of \cite{Kravtsov_2011} to our SRBM ensemble, we arrive at
\begin{equation}
\label{R01}
\langle R_0^{(1)} \rangle=-\frac{\sqrt{2\pi}}{N}\sum_{i\neq j}^{N}e^{-2b_{ij}t^2}2b_{ij}|t|I_0(2b_{ij}t^2),
\end{equation}
where $b_{ij}=\frac{1}{2}\langle|H_{ij}|^2\rangle \simeq \frac{b^2}{2(|i-j|\ln|i-j|)^2}$ and $I_0(x) $ is the 0th order modified Bessel function. In the large $N$ limit, we can replace the sum in Eq.~\eqref{R01} by an integral. Since $\ln \langle R_0\rangle\simeq \langle R_0^{(1)}\rangle$, a direct calculation yields
$-\partial \langle R_0^{(1)}\rangle/\partial \ln \ln t\underrel{t\rightarrow\infty}{\simeq}2b$, which coincides with the value of $d_2$ calculated above for SRBM. Hence,
$\langle R_0\rangle\sim (\ln t)^{-d_2}$,
which indicates an algebraic decay of $\langle R_0\rangle$ in $\ln t$ controlled by the log-multifractal dimension $d_2$. %A similar procedure can be applied to the unitary class with $\beta=2$ as well, see~\cite{LongPaper}. 
On the other hand, for a finite size $N$, the limit $t \rightarrow\infty$ gives $\langle R_0\rangle \sim \langle P_2\rangle \sim (\ln N)^{-d_2}$ (see Eq.~(\ref{eq:P2SMKR})). Therefore, there must exist a characteristic time scale $t^{*}$ separating the infinite-size behavior $\sim (\ln t)^{-d_2}$ of $\langle R_0\rangle$ from its finite-size stationary value $\sim (\ln N)^{-d_2}$, with $ \ln t^*\sim \ln N $.  We can hence assume, as was done in \cite{PhysRevE.108.054127}, the following scaling behavior:
\begin{equation}\label{eq:scaR0}
\begin{split}
    \langle R_0(t,N)\rangle =(\ln N)^{-d_2}g(\ln t/\ln N),
\end{split} 
\end{equation}
with $g(x)\sim_{x\ll 1}x^{-d_2}$ and $g(x)\sim_{x\gg 1} \text{cst}$.
The same scaling behavior is expected for SRUM ensemble, as the amplitudes of its off-diagonal elements decay in the same way. The results, shown in Fig.~\ref{FigDynamics}(a) for SRUM, confirm the validity of the scaling described by Eq.~\eqref{eq:scaR0}.

%\GJ{as of Tuesday}

It is also interesting to examine the spatial expansion of a time-evolving wave packet. In the long-time limit or for sufficiently small $r$, we find that the wave packet amplitudes exhibit a decay behavior similar to the average spatial correlation function of the eigenstates, $C(r)$.
In the short-time limit or for sufficiently large $r$, the wave packet amplitudes decay following, instead, the direct long-range couplings described by the off-diagonal matrix elements, i.e., Eq.~\eqref{eq3}. If we introduce $r_c$ as the crossover scale between the two regimes, these two behaviors can be summarized as 
\begin{equation}
\label{eqpsi2}
\begin{split}
\begin{aligned}
\langle|\psi(r,t)|^{2}\rangle =     \begin{cases}
       & \langle R_0\rangle\left[r(\ln r)^{\alpha}\right]^{-1} , \quad 1< r< r_c, \\
       &B\left[\frac{r}{r_c}\ln(\frac{r}{r_c})\right]^{-2}, \quad r_c< r\leq \frac{N}{2} ,
    \end{cases} 
    \end{aligned}
\end{split}
\end{equation}
with $B= \langle R_0\rangle\left[r_c(\ln r_c)^{\alpha}\right]^{-1}$ \cite{PhysRevE.108.054127, PhysRevLett.79.1959, CHALKER1990253, PhysRevA.100.043612, PhysRevE.86.021136}. %\og{Mention the analogy with the two slopes $\omega^{-\alpha}$ and $\omega^{-2}$ in Kravtsov}. 
%By imposing the normalization condition of the wave packet, we can show, in the same way as in \cite{PhysRevE.108.054127}, that the crossover scale $r_c$ behaves as $r_c \sim t^{d_2/(1-\alpha)}$. Results displayed in Fig.~\ref{FigDynamics}(b) and Fig.~S4 of the Supplemental material \cite{sup} validate the above predictions. 

\emph{Generalization to critical localization.\textemdash} We can construct a whole family of SRBM ensembles if we replace the logarithmic function in Eq.~\eqref{defSRBM} by $\ln^{1+\mu}$ with $\mu\geq 0$.
The parameter $\mu$ influences various properties and behaviors of the system. If we focus on the case of $q=2$ only, analytical procedures as above for $\mu > 0$  lead to $\langle P_2\rangle\sim(\ln N)^{-\mu}+ P_2^{N=\infty}$, with $P_2^{N=\infty}>0$ indicative of a localization behavior. Our numerical observations also reveal that the spatial correlation function of the eigenstates decays as $C(r)\sim 1/\left[r(\ln r)^{1+\mu}\right]$ (see Fig.~S3 of the Supplementary Material \cite{sup}). Noteworthy is that when $\mu=\frac12$ both of these results align with the analytical predictions for %outcomes derived from a supersymmetric nonlinear $\sigma$ model approach applied to
the Anderson model on random regular and Erdős-Rényi graphs \cite{PhysRevLett.72.526,PhysRevB.99.024202,PhysRevLett.67.2049}. This suggests a potential connection between the free parameter $\mu$ and the specific characteristics of certain types of graphs.% However, the precise and rigorous relationship between them remains unclear and warrants further investigation. These results will be discussed more thoroughly in \cite{LongPaper} \cwt{(Sup. Mat. or Long Paper?)}. 

\emph{Conclusion.\textemdash}
In this Letter, we have presented, both analytically and computationally, compelling evidence of log-multifractality at the Anderson transition of effective infinite-dimensionality, through the scaling behavior $\langle P_q \rangle \sim (\ln N)^{-d_q(q-1)}$ for $q > \frac12$.  This scaling behavior signifies a remarkable scale invariance in the logarithm of the system size, extending beyond conventional multifractality in finite dimension and scale invariance in second-order phase transitions. Logarithmic multifractality controls the slow decay of spatial correlations and the slow dynamics of a time-evolving wave packet, particularly its return probability $\langle R_0 \rangle \sim (\ln t)^{-d_2}$. Finally, we discussed how to generalize our random matrix models to obtain other ``critical localization" scenarios predicted at the Anderson transition points on random regular and Erdős-Rényi graphs \cite{PhysRevLett.67.2049,PhysRevLett.72.526,PhysRevB.99.024202}. This work thus uncovers distinctive characteristics of the Anderson transition in infinite dimensionality, demonstrating the existence of transition classes beyond those observed so far \cite{PhysRevB.106.214202,vanoni2023renormalization}. Extensions of this work shall offer opportunities to address related challenges, including strong finite-size effects and slow dynamics near the many-body localization transition.

\acknowledgements
This study was supported by
research funding Grants No.~ANR-18-CE30-0017 and ANR-19-CE30-0013, and by the Singapore
Ministry of Education Academic Research Fund Tier I (WBS
No.~R-144-000-437-114). We thank Calcul en Midi-Pyrénées
(CALMIP) and the National Supercomputing Centre (NSCC) of Singapore for computational resources and assistance. W.~Chen is supported by the President's Graduate Fellowship at National University of Singapore and the Merlion Ph.D. Scholarship awarded by the French Embassy in Singapore.

%\og{(GL: add about the Anderson transition in finite dimension that usually one defines multifractality with respect to $L$, linear size, and not $N$ volume $P_q \sim L^{-\tilde{D}_q(q-1)}$. When the dimension increases, observed numerically at the Anderson transition that $\tilde{D}_q$ decreases (example $\tilde{D}_2=1.4?$ in 3D and $\tilde{D}_2=1.1??$ in 4D. Our log-multifractality corresponds to the limit of infinite dimensionality of $\tilde{D}_q > 0$, as $L \sim \ln N$ in infinite dimension. On the other hand, critical localization corresponds to $\tilde{D}_q = 0$. The question remains whether RRG, with critical localization, is truely representative of the AT$^\infty$ from that perspective.)}

%\nocite{*}
\bibliography{modified_apssamp}% Produces the bibliography via BibTeX.

\providecommand{\noopsort}[1]{}\providecommand{\singleletter}[1]{#1}%
\begin{thebibliography}{106}%
\makeatletter
\providecommand \@ifxundefined [1]{%
 \@ifx{#1\undefined}
}%
\providecommand \@ifnum [1]{%
 \ifnum #1\expandafter \@firstoftwo
 \else \expandafter \@secondoftwo
 \fi
}%
\providecommand \@ifx [1]{%
 \ifx #1\expandafter \@firstoftwo
 \else \expandafter \@secondoftwo
 \fi
}%
\providecommand \natexlab [1]{#1}%
\providecommand \enquote  [1]{``#1''}%
\providecommand \bibnamefont  [1]{#1}%
\providecommand \bibfnamefont [1]{#1}%
\providecommand \citenamefont [1]{#1}%
\providecommand \href@noop [0]{\@secondoftwo}%
\providecommand \href [0]{\begingroup \@sanitize@url \@href}%
\providecommand \@href[1]{\@@startlink{#1}\@@href}%
\providecommand \@@href[1]{\endgroup#1\@@endlink}%
\providecommand \@sanitize@url [0]{\catcode `\\12\catcode `\$12\catcode
  `\&12\catcode `\#12\catcode `\^12\catcode `\_12\catcode `\%12\relax}%
\providecommand \@@startlink[1]{}%
\providecommand \@@endlink[0]{}%
\providecommand \url  [0]{\begingroup\@sanitize@url \@url }%
\providecommand \@url [1]{\endgroup\@href {#1}{\urlprefix }}%
\providecommand \urlprefix  [0]{URL }%
\providecommand \Eprint [0]{\href }%
\providecommand \doibase [0]{https://doi.org/}%
\providecommand \selectlanguage [0]{\@gobble}%
\providecommand \bibinfo  [0]{\@secondoftwo}%
\providecommand \bibfield  [0]{\@secondoftwo}%
\providecommand \translation [1]{[#1]}%
\providecommand \BibitemOpen [0]{}%
\providecommand \bibitemStop [0]{}%
\providecommand \bibitemNoStop [0]{.\EOS\space}%
\providecommand \EOS [0]{\spacefactor3000\relax}%
\providecommand \BibitemShut  [1]{\csname bibitem#1\endcsname}%
\let\auto@bib@innerbib\@empty
%</preamble>
\bibitem [{\citenamefont {Wilson}(1983)}]{wilson1983renormalization}%
  \BibitemOpen
  \bibfield  {author} {\bibinfo {author} {\bibfnamefont {K.~G.}\ \bibnamefont
  {Wilson}},\ }\bibfield  {title} {\bibinfo {title} {{T}he renormalization
  group and critical phenomena},\ }\href
  {https://doi.org/10.1103/RevModPhys.55.583} {\bibfield  {journal} {\bibinfo
  {journal} {Rev. Mod. Phys.}\ }\textbf {\bibinfo {volume} {55}},\ \bibinfo
  {pages} {583} (\bibinfo {year} {1983})}\BibitemShut {NoStop}%
\bibitem [{\citenamefont {Kosterlitz}(2016)}]{kosterlitz2016kosterlitz}%
  \BibitemOpen
  \bibfield  {author} {\bibinfo {author} {\bibfnamefont {J.~M.}\ \bibnamefont
  {Kosterlitz}},\ }\bibfield  {title} {\bibinfo {title}
  {{K}osterlitz--{T}houless physics: a review of key issues},\ }\href
  {https://dx.doi.org/10.1088/0034-4885/79/2/026001} {\bibfield  {journal}
  {\bibinfo  {journal} {Reports on Progress in Physics}\ }\textbf {\bibinfo
  {volume} {79}},\ \bibinfo {pages} {026001} (\bibinfo {year}
  {2016})}\BibitemShut {NoStop}%
\bibitem [{\citenamefont {Hsieh}\ \emph {et~al.}(2013)\citenamefont {Hsieh},
  \citenamefont {Kao},\ and\ \citenamefont {Sandvik}}]{hsieh2013finite}%
  \BibitemOpen
  \bibfield  {author} {\bibinfo {author} {\bibfnamefont {Y.-D.}\ \bibnamefont
  {Hsieh}}, \bibinfo {author} {\bibfnamefont {Y.-J.}\ \bibnamefont {Kao}},\
  and\ \bibinfo {author} {\bibfnamefont {A.~W.}\ \bibnamefont {Sandvik}},\
  }\bibfield  {title} {\bibinfo {title} {{F}inite-size scaling method for the
  {B}erezinskii--{K}osterlitz--{T}houless transition},\ }\href
  {https://dx.doi.org/10.1088/1742-5468/2013/09/P09001} {\bibfield  {journal}
  {\bibinfo  {journal} {Journal of Statistical Mechanics: Theory and
  Experiment}\ }\textbf {\bibinfo {volume} {2013}},\ \bibinfo {pages} {P09001}
  (\bibinfo {year} {2013})}\BibitemShut {NoStop}%
\bibitem [{\citenamefont {Evers}\ and\ \citenamefont
  {Mirlin}(2008)}]{RevModPhys.80.1355}%
  \BibitemOpen
  \bibfield  {author} {\bibinfo {author} {\bibfnamefont {F.}~\bibnamefont
  {Evers}}\ and\ \bibinfo {author} {\bibfnamefont {A.~D.}\ \bibnamefont
  {Mirlin}},\ }\bibfield  {title} {\bibinfo {title} {{A}nderson transitions},\
  }\href {https://doi.org/10.1103/RevModPhys.80.1355} {\bibfield  {journal}
  {\bibinfo  {journal} {Rev. Mod. Phys.}\ }\textbf {\bibinfo {volume} {80}},\
  \bibinfo {pages} {1355} (\bibinfo {year} {2008})}\BibitemShut {NoStop}%
\bibitem [{\citenamefont {Abrahams}(2010)}]{abrahams201050}%
  \BibitemOpen
  \bibfield  {author} {\bibinfo {author} {\bibfnamefont {E.}~\bibnamefont
  {Abrahams}},\ }\href@noop {} {\emph {\bibinfo {title} {50 years of {A}nderson
  {L}ocalization}}},\ Vol.~\bibinfo {volume} {24}\ (\bibinfo  {publisher}
  {world scientific},\ \bibinfo {year} {2010})\BibitemShut {NoStop}%
\bibitem [{\citenamefont {Castellani}\ and\ \citenamefont
  {Peliti}(1986)}]{castellani1986multifractal}%
  \BibitemOpen
  \bibfield  {author} {\bibinfo {author} {\bibfnamefont {C.}~\bibnamefont
  {Castellani}}\ and\ \bibinfo {author} {\bibfnamefont {L.}~\bibnamefont
  {Peliti}},\ }\bibfield  {title} {\bibinfo {title} {{M}ultifractal
  wavefunction at the localisation threshold},\ }\href
  {https://dx.doi.org/10.1088/0305-4470/19/8/004} {\bibfield  {journal}
  {\bibinfo  {journal} {Journal of physics A: mathematical and general}\
  }\textbf {\bibinfo {volume} {19}},\ \bibinfo {pages} {L429} (\bibinfo {year}
  {1986})}\BibitemShut {NoStop}%
\bibitem [{\citenamefont {Rodriguez}\ \emph {et~al.}(2009)\citenamefont
  {Rodriguez}, \citenamefont {Vasquez},\ and\ \citenamefont
  {R\"omer}}]{PhysRevLett.102.106406}%
  \BibitemOpen
  \bibfield  {author} {\bibinfo {author} {\bibfnamefont {A.}~\bibnamefont
  {Rodriguez}}, \bibinfo {author} {\bibfnamefont {L.~J.}\ \bibnamefont
  {Vasquez}},\ and\ \bibinfo {author} {\bibfnamefont {R.~A.}\ \bibnamefont
  {R\"omer}},\ }\bibfield  {title} {\bibinfo {title} {{M}ultifractal {A}nalysis
  with the {P}robability {D}ensity {F}unction at the {T}hree-{D}imensional
  {A}nderson {T}ransition},\ }\href
  {https://doi.org/10.1103/PhysRevLett.102.106406} {\bibfield  {journal}
  {\bibinfo  {journal} {Phys. Rev. Lett.}\ }\textbf {\bibinfo {volume} {102}},\
  \bibinfo {pages} {106406} (\bibinfo {year} {2009})}\BibitemShut {NoStop}%
\bibitem [{\citenamefont {Faez}\ \emph {et~al.}(2009)\citenamefont {Faez},
  \citenamefont {Strybulevych}, \citenamefont {Page}, \citenamefont
  {Lagendijk},\ and\ \citenamefont {van Tiggelen}}]{PhysRevLett.103.155703}%
  \BibitemOpen
  \bibfield  {author} {\bibinfo {author} {\bibfnamefont {S.}~\bibnamefont
  {Faez}}, \bibinfo {author} {\bibfnamefont {A.}~\bibnamefont {Strybulevych}},
  \bibinfo {author} {\bibfnamefont {J.~H.}\ \bibnamefont {Page}}, \bibinfo
  {author} {\bibfnamefont {A.}~\bibnamefont {Lagendijk}},\ and\ \bibinfo
  {author} {\bibfnamefont {B.~A.}\ \bibnamefont {van Tiggelen}},\ }\bibfield
  {title} {\bibinfo {title} {{O}bservation of {M}ultifractality in {A}nderson
  {L}ocalization of {U}ltrasound},\ }\href
  {https://doi.org/10.1103/PhysRevLett.103.155703} {\bibfield  {journal}
  {\bibinfo  {journal} {Phys. Rev. Lett.}\ }\textbf {\bibinfo {volume} {103}},\
  \bibinfo {pages} {155703} (\bibinfo {year} {2009})}\BibitemShut {NoStop}%
\bibitem [{\citenamefont {Rodriguez}\ \emph {et~al.}(2011)\citenamefont
  {Rodriguez}, \citenamefont {Vasquez}, \citenamefont {Slevin},\ and\
  \citenamefont {R\"omer}}]{PhysRevB.84.134209}%
  \BibitemOpen
  \bibfield  {author} {\bibinfo {author} {\bibfnamefont {A.}~\bibnamefont
  {Rodriguez}}, \bibinfo {author} {\bibfnamefont {L.~J.}\ \bibnamefont
  {Vasquez}}, \bibinfo {author} {\bibfnamefont {K.}~\bibnamefont {Slevin}},\
  and\ \bibinfo {author} {\bibfnamefont {R.~A.}\ \bibnamefont {R\"omer}},\
  }\bibfield  {title} {\bibinfo {title} {{M}ultifractal finite-size scaling and
  universality at the {A}nderson transition},\ }\href
  {https://doi.org/10.1103/PhysRevB.84.134209} {\bibfield  {journal} {\bibinfo
  {journal} {Phys. Rev. B}\ }\textbf {\bibinfo {volume} {84}},\ \bibinfo
  {pages} {134209} (\bibinfo {year} {2011})}\BibitemShut {NoStop}%
\bibitem [{\citenamefont {Lemari\'e}\ \emph {et~al.}(2009)\citenamefont
  {Lemari\'e}, \citenamefont {Chab\'e}, \citenamefont {Szriftgiser},
  \citenamefont {Garreau}, \citenamefont {Gr\'emaud},\ and\ \citenamefont
  {Delande}}]{PhysRevA.80.043626}%
  \BibitemOpen
  \bibfield  {author} {\bibinfo {author} {\bibfnamefont {G.}~\bibnamefont
  {Lemari\'e}}, \bibinfo {author} {\bibfnamefont {J.}~\bibnamefont {Chab\'e}},
  \bibinfo {author} {\bibfnamefont {P.}~\bibnamefont {Szriftgiser}}, \bibinfo
  {author} {\bibfnamefont {J.~C.}\ \bibnamefont {Garreau}}, \bibinfo {author}
  {\bibfnamefont {B.}~\bibnamefont {Gr\'emaud}},\ and\ \bibinfo {author}
  {\bibfnamefont {D.}~\bibnamefont {Delande}},\ }\bibfield  {title} {\bibinfo
  {title} {{O}bservation of the {A}nderson metal-insulator transition with
  atomic matter waves: {T}heory and experiment},\ }\href
  {https://doi.org/10.1103/PhysRevA.80.043626} {\bibfield  {journal} {\bibinfo
  {journal} {Phys. Rev. A}\ }\textbf {\bibinfo {volume} {80}},\ \bibinfo
  {pages} {043626} (\bibinfo {year} {2009})}\BibitemShut {NoStop}%
\bibitem [{\citenamefont {Lemari\'e}\ \emph {et~al.}(2010)\citenamefont
  {Lemari\'e}, \citenamefont {Lignier}, \citenamefont {Delande}, \citenamefont
  {Szriftgiser},\ and\ \citenamefont {Garreau}}]{PhysRevLett.105.090601}%
  \BibitemOpen
  \bibfield  {author} {\bibinfo {author} {\bibfnamefont {G.}~\bibnamefont
  {Lemari\'e}}, \bibinfo {author} {\bibfnamefont {H.}~\bibnamefont {Lignier}},
  \bibinfo {author} {\bibfnamefont {D.}~\bibnamefont {Delande}}, \bibinfo
  {author} {\bibfnamefont {P.}~\bibnamefont {Szriftgiser}},\ and\ \bibinfo
  {author} {\bibfnamefont {J.~C.}\ \bibnamefont {Garreau}},\ }\bibfield
  {title} {\bibinfo {title} {{C}ritical {S}tate of the {A}nderson {T}ransition:
  {B}etween a {M}etal and an {I}nsulator},\ }\href
  {https://doi.org/10.1103/PhysRevLett.105.090601} {\bibfield  {journal}
  {\bibinfo  {journal} {Phys. Rev. Lett.}\ }\textbf {\bibinfo {volume} {105}},\
  \bibinfo {pages} {090601} (\bibinfo {year} {2010})}\BibitemShut {NoStop}%
\bibitem [{\citenamefont {Chab\'e}\ \emph {et~al.}(2008)\citenamefont
  {Chab\'e}, \citenamefont {Lemari\'e}, \citenamefont {Gr\'emaud},
  \citenamefont {Delande}, \citenamefont {Szriftgiser},\ and\ \citenamefont
  {Garreau}}]{PhysRevLett.101.255702}%
  \BibitemOpen
  \bibfield  {author} {\bibinfo {author} {\bibfnamefont {J.}~\bibnamefont
  {Chab\'e}}, \bibinfo {author} {\bibfnamefont {G.}~\bibnamefont {Lemari\'e}},
  \bibinfo {author} {\bibfnamefont {B.}~\bibnamefont {Gr\'emaud}}, \bibinfo
  {author} {\bibfnamefont {D.}~\bibnamefont {Delande}}, \bibinfo {author}
  {\bibfnamefont {P.}~\bibnamefont {Szriftgiser}},\ and\ \bibinfo {author}
  {\bibfnamefont {J.~C.}\ \bibnamefont {Garreau}},\ }\bibfield  {title}
  {\bibinfo {title} {{E}xperimental {O}bservation of the {A}nderson
  {M}etal-{I}nsulator {T}ransition with {A}tomic {M}atter {W}aves},\ }\href
  {https://doi.org/10.1103/PhysRevLett.101.255702} {\bibfield  {journal}
  {\bibinfo  {journal} {Phys. Rev. Lett.}\ }\textbf {\bibinfo {volume} {101}},\
  \bibinfo {pages} {255702} (\bibinfo {year} {2008})}\BibitemShut {NoStop}%
\bibitem [{\citenamefont {Cuevas}\ \emph {et~al.}(2001)\citenamefont {Cuevas},
  \citenamefont {Gasparian},\ and\ \citenamefont
  {Ortu\~no}}]{PhysRevLett.87.056601}%
  \BibitemOpen
  \bibfield  {author} {\bibinfo {author} {\bibfnamefont {E.}~\bibnamefont
  {Cuevas}}, \bibinfo {author} {\bibfnamefont {V.}~\bibnamefont {Gasparian}},\
  and\ \bibinfo {author} {\bibfnamefont {M.}~\bibnamefont {Ortu\~no}},\
  }\bibfield  {title} {\bibinfo {title} {{A}nomalously {L}arge {C}ritical
  {R}egions in {P}ower-{L}aw {R}andom {M}atrix {E}nsembles},\ }\href
  {https://doi.org/10.1103/PhysRevLett.87.056601} {\bibfield  {journal}
  {\bibinfo  {journal} {Phys. Rev. Lett.}\ }\textbf {\bibinfo {volume} {87}},\
  \bibinfo {pages} {056601} (\bibinfo {year} {2001})}\BibitemShut {NoStop}%
\bibitem [{\citenamefont {Tarquini}\ \emph {et~al.}(2017)\citenamefont
  {Tarquini}, \citenamefont {Biroli},\ and\ \citenamefont
  {Tarzia}}]{PhysRevB.95.094204}%
  \BibitemOpen
  \bibfield  {author} {\bibinfo {author} {\bibfnamefont {E.}~\bibnamefont
  {Tarquini}}, \bibinfo {author} {\bibfnamefont {G.}~\bibnamefont {Biroli}},\
  and\ \bibinfo {author} {\bibfnamefont {M.}~\bibnamefont {Tarzia}},\
  }\bibfield  {title} {\bibinfo {title} {{C}ritical properties of the
  {A}nderson localization transition and the high-dimensional limit},\ }\href
  {https://doi.org/10.1103/PhysRevB.95.094204} {\bibfield  {journal} {\bibinfo
  {journal} {Phys. Rev. B}\ }\textbf {\bibinfo {volume} {95}},\ \bibinfo
  {pages} {094204} (\bibinfo {year} {2017})}\BibitemShut {NoStop}%
\bibitem [{\citenamefont {Ohtsuki}\ and\ \citenamefont
  {Kawarabayashi}(1997)}]{OhtsukiKawarabayashi}%
  \BibitemOpen
  \bibfield  {author} {\bibinfo {author} {\bibfnamefont {T.}~\bibnamefont
  {Ohtsuki}}\ and\ \bibinfo {author} {\bibfnamefont {T.}~\bibnamefont
  {Kawarabayashi}},\ }\bibfield  {title} {\bibinfo {title} {{A}nomalous
  {D}iffusion at the {A}nderson {T}ransitions},\ }\href
  {https://doi.org/10.1143/JPSJ.66.314} {\bibfield  {journal} {\bibinfo
  {journal} {Journal of the Physical Society of Japan}\ }\textbf {\bibinfo
  {volume} {66}},\ \bibinfo {pages} {314} (\bibinfo {year} {1997})}\BibitemShut
  {NoStop}%
\bibitem [{\citenamefont {M\"uller}\ \emph {et~al.}(2016)\citenamefont
  {M\"uller}, \citenamefont {Delande},\ and\ \citenamefont
  {Shapiro}}]{PhysRevA.94.033615}%
  \BibitemOpen
  \bibfield  {author} {\bibinfo {author} {\bibfnamefont {C.~A.}\ \bibnamefont
  {M\"uller}}, \bibinfo {author} {\bibfnamefont {D.}~\bibnamefont {Delande}},\
  and\ \bibinfo {author} {\bibfnamefont {B.}~\bibnamefont {Shapiro}},\
  }\bibfield  {title} {\bibinfo {title} {{C}ritical dynamics at the {A}nderson
  localization mobility edge},\ }\href
  {https://doi.org/10.1103/PhysRevA.94.033615} {\bibfield  {journal} {\bibinfo
  {journal} {Phys. Rev. A}\ }\textbf {\bibinfo {volume} {94}},\ \bibinfo
  {pages} {033615} (\bibinfo {year} {2016})}\BibitemShut {NoStop}%
\bibitem [{\citenamefont {Ketzmerick}\ \emph {et~al.}(1992)\citenamefont
  {Ketzmerick}, \citenamefont {Petschel},\ and\ \citenamefont
  {Geisel}}]{PhysRevLett.69.695}%
  \BibitemOpen
  \bibfield  {author} {\bibinfo {author} {\bibfnamefont {R.}~\bibnamefont
  {Ketzmerick}}, \bibinfo {author} {\bibfnamefont {G.}~\bibnamefont
  {Petschel}},\ and\ \bibinfo {author} {\bibfnamefont {T.}~\bibnamefont
  {Geisel}},\ }\bibfield  {title} {\bibinfo {title} {{S}low decay of temporal
  correlations in quantum systems with {C}antor spectra},\ }\href
  {https://doi.org/10.1103/PhysRevLett.69.695} {\bibfield  {journal} {\bibinfo
  {journal} {Phys. Rev. Lett.}\ }\textbf {\bibinfo {volume} {69}},\ \bibinfo
  {pages} {695} (\bibinfo {year} {1992})}\BibitemShut {NoStop}%
\bibitem [{\citenamefont {Kravtsov}\ \emph {et~al.}(2010)\citenamefont
  {Kravtsov}, \citenamefont {Ossipov}, \citenamefont {Yevtushenko},\ and\
  \citenamefont {Cuevas}}]{PhysRevB.82.161102}%
  \BibitemOpen
  \bibfield  {author} {\bibinfo {author} {\bibfnamefont {V.~E.}\ \bibnamefont
  {Kravtsov}}, \bibinfo {author} {\bibfnamefont {A.}~\bibnamefont {Ossipov}},
  \bibinfo {author} {\bibfnamefont {O.~M.}\ \bibnamefont {Yevtushenko}},\ and\
  \bibinfo {author} {\bibfnamefont {E.}~\bibnamefont {Cuevas}},\ }\bibfield
  {title} {\bibinfo {title} {{D}ynamical scaling for critical states:
  {V}alidity of {C}halker's ansatz for strong fractality},\ }\href
  {https://doi.org/10.1103/PhysRevB.82.161102} {\bibfield  {journal} {\bibinfo
  {journal} {Phys. Rev. B}\ }\textbf {\bibinfo {volume} {82}},\ \bibinfo
  {pages} {161102} (\bibinfo {year} {2010})}\BibitemShut {NoStop}%
\bibitem [{\citenamefont {Kravtsov}\ \emph {et~al.}(2011)\citenamefont
  {Kravtsov}, \citenamefont {Ossipov},\ and\ \citenamefont
  {Yevtushenko}}]{Kravtsov_2011}%
  \BibitemOpen
  \bibfield  {author} {\bibinfo {author} {\bibfnamefont {V.~E.}\ \bibnamefont
  {Kravtsov}}, \bibinfo {author} {\bibfnamefont {A.}~\bibnamefont {Ossipov}},\
  and\ \bibinfo {author} {\bibfnamefont {O.~M.}\ \bibnamefont {Yevtushenko}},\
  }\bibfield  {title} {\bibinfo {title} {{R}eturn probability and scaling
  exponents in the critical random matrix ensemble},\ }\href
  {https://doi.org/10.1088/1751-8113/44/30/305003} {\bibfield  {journal}
  {\bibinfo  {journal} {Journal of Physics A: Mathematical and Theoretical}\
  }\textbf {\bibinfo {volume} {44}},\ \bibinfo {pages} {305003} (\bibinfo
  {year} {2011})}\BibitemShut {NoStop}%
\bibitem [{\citenamefont {Altshuler}\ and\ \citenamefont
  {Kravtsov}(2023)}]{ALTSHULER2023169300}%
  \BibitemOpen
  \bibfield  {author} {\bibinfo {author} {\bibfnamefont {B.}~\bibnamefont
  {Altshuler}}\ and\ \bibinfo {author} {\bibfnamefont {V.}~\bibnamefont
  {Kravtsov}},\ }\bibfield  {title} {\bibinfo {title} {{R}andom {C}antor sets
  and mini-bands in local spectrum of quantum systems},\ }\href
  {https://doi.org/https://doi.org/10.1016/j.aop.2023.169300} {\bibfield
  {journal} {\bibinfo  {journal} {Annals of Physics}\ }\textbf {\bibinfo
  {volume} {456}},\ \bibinfo {pages} {169300} (\bibinfo {year}
  {2023})}\BibitemShut {NoStop}%
\bibitem [{\citenamefont {Ghosh}\ \emph {et~al.}(2017)\citenamefont {Ghosh},
  \citenamefont {Miniatura}, \citenamefont {Cherroret},\ and\ \citenamefont
  {Delande}}]{PhysRevA.95.041602}%
  \BibitemOpen
  \bibfield  {author} {\bibinfo {author} {\bibfnamefont {S.}~\bibnamefont
  {Ghosh}}, \bibinfo {author} {\bibfnamefont {C.}~\bibnamefont {Miniatura}},
  \bibinfo {author} {\bibfnamefont {N.}~\bibnamefont {Cherroret}},\ and\
  \bibinfo {author} {\bibfnamefont {D.}~\bibnamefont {Delande}},\ }\bibfield
  {title} {\bibinfo {title} {{C}oherent forward scattering as a signature of
  {A}nderson metal-insulator transitions},\ }\href
  {https://doi.org/10.1103/PhysRevA.95.041602} {\bibfield  {journal} {\bibinfo
  {journal} {Phys. Rev. A}\ }\textbf {\bibinfo {volume} {95}},\ \bibinfo
  {pages} {041602} (\bibinfo {year} {2017})}\BibitemShut {NoStop}%
\bibitem [{\citenamefont {Martinez}\ \emph {et~al.}(2023)\citenamefont
  {Martinez}, \citenamefont {Lemarié}, \citenamefont {Georgeot}, \citenamefont
  {Miniatura},\ and\ \citenamefont {Giraud}}]{martinez2022coherent}%
  \BibitemOpen
  \bibfield  {author} {\bibinfo {author} {\bibfnamefont {M.}~\bibnamefont
  {Martinez}}, \bibinfo {author} {\bibfnamefont {G.}~\bibnamefont {Lemarié}},
  \bibinfo {author} {\bibfnamefont {B.}~\bibnamefont {Georgeot}}, \bibinfo
  {author} {\bibfnamefont {C.}~\bibnamefont {Miniatura}},\ and\ \bibinfo
  {author} {\bibfnamefont {O.}~\bibnamefont {Giraud}},\ }\bibfield  {title}
  {\bibinfo {title} {{Coherent forward scattering as a robust probe of
  multifractality in critical disordered media}},\ }\href
  {https://doi.org/10.21468/SciPostPhys.14.3.057} {\bibfield  {journal}
  {\bibinfo  {journal} {SciPost Phys.}\ }\textbf {\bibinfo {volume} {14}},\
  \bibinfo {pages} {057} (\bibinfo {year} {2023})}\BibitemShut {NoStop}%
\bibitem [{\citenamefont {Akridas-Morel}\ \emph {et~al.}(2019)\citenamefont
  {Akridas-Morel}, \citenamefont {Cherroret},\ and\ \citenamefont
  {Delande}}]{PhysRevA.100.043612}%
  \BibitemOpen
  \bibfield  {author} {\bibinfo {author} {\bibfnamefont {P.}~\bibnamefont
  {Akridas-Morel}}, \bibinfo {author} {\bibfnamefont {N.}~\bibnamefont
  {Cherroret}},\ and\ \bibinfo {author} {\bibfnamefont {D.}~\bibnamefont
  {Delande}},\ }\bibfield  {title} {\bibinfo {title} {{M}ultifractality of the
  kicked rotor at the critical point of the {A}nderson transition},\ }\href
  {https://doi.org/10.1103/PhysRevA.100.043612} {\bibfield  {journal} {\bibinfo
   {journal} {Phys. Rev. A}\ }\textbf {\bibinfo {volume} {100}},\ \bibinfo
  {pages} {043612} (\bibinfo {year} {2019})}\BibitemShut {NoStop}%
\bibitem [{\citenamefont {Chen}\ \emph
  {et~al.}(2023{\natexlab{a}})\citenamefont {Chen}, \citenamefont {Lemari\'e},\
  and\ \citenamefont {Gong}}]{PhysRevE.108.054127}%
  \BibitemOpen
  \bibfield  {author} {\bibinfo {author} {\bibfnamefont {W.}~\bibnamefont
  {Chen}}, \bibinfo {author} {\bibfnamefont {G.}~\bibnamefont {Lemari\'e}},\
  and\ \bibinfo {author} {\bibfnamefont {J.}~\bibnamefont {Gong}},\ }\bibfield
  {title} {\bibinfo {title} {{C}ritical dynamics of long-range quantum
  disordered systems},\ }\href {https://doi.org/10.1103/PhysRevE.108.054127}
  {\bibfield  {journal} {\bibinfo  {journal} {Phys. Rev. E}\ }\textbf {\bibinfo
  {volume} {108}},\ \bibinfo {pages} {054127} (\bibinfo {year}
  {2023}{\natexlab{a}})}\BibitemShut {NoStop}%
\bibitem [{\citenamefont {Mandelbrot}(1974)}]{mandelbrot1974intermittent}%
  \BibitemOpen
  \bibfield  {author} {\bibinfo {author} {\bibfnamefont {B.~B.}\ \bibnamefont
  {Mandelbrot}},\ }\bibfield  {title} {\bibinfo {title} {{I}ntermittent
  turbulence in self-similar cascades: divergence of high moments and dimension
  of the carrier},\ }\href@noop {} {\bibfield  {journal} {\bibinfo  {journal}
  {Journal of fluid Mechanics}\ }\textbf {\bibinfo {volume} {62}},\ \bibinfo
  {pages} {331} (\bibinfo {year} {1974})}\BibitemShut {NoStop}%
\bibitem [{\citenamefont {Mandelbrot}\ and\ \citenamefont
  {Mandelbrot}(1982)}]{mandelbrot1982fractal}%
  \BibitemOpen
  \bibfield  {author} {\bibinfo {author} {\bibfnamefont {B.~B.}\ \bibnamefont
  {Mandelbrot}}\ and\ \bibinfo {author} {\bibfnamefont {B.~B.}\ \bibnamefont
  {Mandelbrot}},\ }\href@noop {} {\emph {\bibinfo {title} {{T}he fractal
  geometry of nature}}},\ Vol.~\bibinfo {volume} {1}\ (\bibinfo  {publisher}
  {WH freeman New York},\ \bibinfo {year} {1982})\BibitemShut {NoStop}%
\bibitem [{\citenamefont {Falconer}(2004)}]{falconer2004fractal}%
  \BibitemOpen
  \bibfield  {author} {\bibinfo {author} {\bibfnamefont {K.}~\bibnamefont
  {Falconer}},\ }\href@noop {} {\emph {\bibinfo {title} {{F}ractal geometry:
  mathematical foundations and applications}}}\ (\bibinfo  {publisher} {John
  Wiley \& Sons},\ \bibinfo {year} {2004})\BibitemShut {NoStop}%
\bibitem [{\citenamefont {Zirnbauer}(1986{\natexlab{a}})}]{PhysRevB.34.6394}%
  \BibitemOpen
  \bibfield  {author} {\bibinfo {author} {\bibfnamefont {M.~R.}\ \bibnamefont
  {Zirnbauer}},\ }\bibfield  {title} {\bibinfo {title} {{L}ocalization
  transition on the {B}ethe lattice},\ }\href
  {https://doi.org/10.1103/PhysRevB.34.6394} {\bibfield  {journal} {\bibinfo
  {journal} {Phys. Rev. B}\ }\textbf {\bibinfo {volume} {34}},\ \bibinfo
  {pages} {6394} (\bibinfo {year} {1986}{\natexlab{a}})}\BibitemShut {NoStop}%
\bibitem [{\citenamefont {Zirnbauer}(1986{\natexlab{b}})}]{ZIRNBAUER1986375}%
  \BibitemOpen
  \bibfield  {author} {\bibinfo {author} {\bibfnamefont {M.~R.}\ \bibnamefont
  {Zirnbauer}},\ }\bibfield  {title} {\bibinfo {title} {{A}nderson localization
  and non-linear sigma model with graded symmetry},\ }\href
  {https://doi.org/https://doi.org/10.1016/0550-3213(86)90316-0} {\bibfield
  {journal} {\bibinfo  {journal} {Nuclear Physics B}\ }\textbf {\bibinfo
  {volume} {265}},\ \bibinfo {pages} {375} (\bibinfo {year}
  {1986}{\natexlab{b}})}\BibitemShut {NoStop}%
\bibitem [{\citenamefont {Fyodorov}\ and\ \citenamefont
  {Mirlin}(1991)}]{PhysRevLett.67.2049}%
  \BibitemOpen
  \bibfield  {author} {\bibinfo {author} {\bibfnamefont {Y.~V.}\ \bibnamefont
  {Fyodorov}}\ and\ \bibinfo {author} {\bibfnamefont {A.~D.}\ \bibnamefont
  {Mirlin}},\ }\bibfield  {title} {\bibinfo {title} {{L}ocalization in ensemble
  of sparse random matrices},\ }\href
  {https://doi.org/10.1103/PhysRevLett.67.2049} {\bibfield  {journal} {\bibinfo
   {journal} {Phys. Rev. Lett.}\ }\textbf {\bibinfo {volume} {67}},\ \bibinfo
  {pages} {2049} (\bibinfo {year} {1991})}\BibitemShut {NoStop}%
\bibitem [{\citenamefont {Mirlin}\ and\ \citenamefont
  {Fyodorov}(1994)}]{PhysRevLett.72.526}%
  \BibitemOpen
  \bibfield  {author} {\bibinfo {author} {\bibfnamefont {A.~D.}\ \bibnamefont
  {Mirlin}}\ and\ \bibinfo {author} {\bibfnamefont {Y.~V.}\ \bibnamefont
  {Fyodorov}},\ }\bibfield  {title} {\bibinfo {title} {{D}istribution of local
  densities of states, order parameter function, and critical behavior near the
  {A}nderson transition},\ }\href {https://doi.org/10.1103/PhysRevLett.72.526}
  {\bibfield  {journal} {\bibinfo  {journal} {Phys. Rev. Lett.}\ }\textbf
  {\bibinfo {volume} {72}},\ \bibinfo {pages} {526} (\bibinfo {year}
  {1994})}\BibitemShut {NoStop}%
\bibitem [{\citenamefont {Monthus}\ and\ \citenamefont
  {Garel}(2011)}]{Monthus_2011}%
  \BibitemOpen
  \bibfield  {author} {\bibinfo {author} {\bibfnamefont {C.}~\bibnamefont
  {Monthus}}\ and\ \bibinfo {author} {\bibfnamefont {T.}~\bibnamefont
  {Garel}},\ }\bibfield  {title} {\bibinfo {title} {{A}nderson localization on
  the {C}ayley tree: multifractal statistics of the transmission at criticality
  and off criticality},\ }\href
  {https://doi.org/10.1088/1751-8113/44/14/145001} {\bibfield  {journal}
  {\bibinfo  {journal} {Journal of Physics A: Mathematical and Theoretical}\
  }\textbf {\bibinfo {volume} {44}},\ \bibinfo {pages} {145001} (\bibinfo
  {year} {2011})}\BibitemShut {NoStop}%
\bibitem [{\citenamefont {Biroli}\ \emph {et~al.}(2012)\citenamefont {Biroli},
  \citenamefont {Ribeiro-Teixeira},\ and\ \citenamefont
  {Tarzia}}]{biroli2012difference}%
  \BibitemOpen
  \bibfield  {author} {\bibinfo {author} {\bibfnamefont {G.}~\bibnamefont
  {Biroli}}, \bibinfo {author} {\bibfnamefont {A.~C.}\ \bibnamefont
  {Ribeiro-Teixeira}},\ and\ \bibinfo {author} {\bibfnamefont {M.}~\bibnamefont
  {Tarzia}},\ }\href@noop {} {\bibinfo {title} {{D}ifference between level
  statistics, ergodicity and localization transitions on the {B}ethe lattice}}
  (\bibinfo {year} {2012}),\ \Eprint {https://arxiv.org/abs/1211.7334}
  {arXiv:1211.7334 [cond-mat.dis-nn]} \BibitemShut {NoStop}%
\bibitem [{\citenamefont {De~Luca}\ \emph {et~al.}(2014)\citenamefont
  {De~Luca}, \citenamefont {Altshuler}, \citenamefont {Kravtsov},\ and\
  \citenamefont {Scardicchio}}]{PhysRevLett.113.046806}%
  \BibitemOpen
  \bibfield  {author} {\bibinfo {author} {\bibfnamefont {A.}~\bibnamefont
  {De~Luca}}, \bibinfo {author} {\bibfnamefont {B.~L.}\ \bibnamefont
  {Altshuler}}, \bibinfo {author} {\bibfnamefont {V.~E.}\ \bibnamefont
  {Kravtsov}},\ and\ \bibinfo {author} {\bibfnamefont {A.}~\bibnamefont
  {Scardicchio}},\ }\bibfield  {title} {\bibinfo {title} {{A}nderson
  {L}ocalization on the {B}ethe {L}attice: {N}onergodicity of {E}xtended
  {S}tates},\ }\href {https://doi.org/10.1103/PhysRevLett.113.046806}
  {\bibfield  {journal} {\bibinfo  {journal} {Phys. Rev. Lett.}\ }\textbf
  {\bibinfo {volume} {113}},\ \bibinfo {pages} {046806} (\bibinfo {year}
  {2014})}\BibitemShut {NoStop}%
\bibitem [{\citenamefont {Altshuler}\ \emph {et~al.}(2016)\citenamefont
  {Altshuler}, \citenamefont {Cuevas}, \citenamefont {Ioffe},\ and\
  \citenamefont {Kravtsov}}]{PhysRevLett.117.156601}%
  \BibitemOpen
  \bibfield  {author} {\bibinfo {author} {\bibfnamefont {B.~L.}\ \bibnamefont
  {Altshuler}}, \bibinfo {author} {\bibfnamefont {E.}~\bibnamefont {Cuevas}},
  \bibinfo {author} {\bibfnamefont {L.~B.}\ \bibnamefont {Ioffe}},\ and\
  \bibinfo {author} {\bibfnamefont {V.~E.}\ \bibnamefont {Kravtsov}},\
  }\bibfield  {title} {\bibinfo {title} {{N}onergodic {P}hases in {S}trongly
  {D}isordered {R}andom {R}egular {G}raphs},\ }\href
  {https://doi.org/10.1103/PhysRevLett.117.156601} {\bibfield  {journal}
  {\bibinfo  {journal} {Phys. Rev. Lett.}\ }\textbf {\bibinfo {volume} {117}},\
  \bibinfo {pages} {156601} (\bibinfo {year} {2016})}\BibitemShut {NoStop}%
\bibitem [{\citenamefont {Tikhonov}\ \emph {et~al.}(2016)\citenamefont
  {Tikhonov}, \citenamefont {Mirlin},\ and\ \citenamefont
  {Skvortsov}}]{PhysRevB.94.220203}%
  \BibitemOpen
  \bibfield  {author} {\bibinfo {author} {\bibfnamefont {K.~S.}\ \bibnamefont
  {Tikhonov}}, \bibinfo {author} {\bibfnamefont {A.~D.}\ \bibnamefont
  {Mirlin}},\ and\ \bibinfo {author} {\bibfnamefont {M.~A.}\ \bibnamefont
  {Skvortsov}},\ }\bibfield  {title} {\bibinfo {title} {{A}nderson localization
  and ergodicity on random regular graphs},\ }\href
  {https://doi.org/10.1103/PhysRevB.94.220203} {\bibfield  {journal} {\bibinfo
  {journal} {Phys. Rev. B}\ }\textbf {\bibinfo {volume} {94}},\ \bibinfo
  {pages} {220203} (\bibinfo {year} {2016})}\BibitemShut {NoStop}%
\bibitem [{\citenamefont {Tikhonov}\ and\ \citenamefont
  {Mirlin}(2016)}]{PhysRevB.94.184203}%
  \BibitemOpen
  \bibfield  {author} {\bibinfo {author} {\bibfnamefont {K.~S.}\ \bibnamefont
  {Tikhonov}}\ and\ \bibinfo {author} {\bibfnamefont {A.~D.}\ \bibnamefont
  {Mirlin}},\ }\bibfield  {title} {\bibinfo {title} {{F}ractality of wave
  functions on a {C}ayley tree: {D}ifference between tree and locally treelike
  graph without boundary},\ }\href {https://doi.org/10.1103/PhysRevB.94.184203}
  {\bibfield  {journal} {\bibinfo  {journal} {Phys. Rev. B}\ }\textbf {\bibinfo
  {volume} {94}},\ \bibinfo {pages} {184203} (\bibinfo {year}
  {2016})}\BibitemShut {NoStop}%
\bibitem [{\citenamefont {Sonner}\ \emph {et~al.}(2017)\citenamefont {Sonner},
  \citenamefont {Tikhonov},\ and\ \citenamefont {Mirlin}}]{PhysRevB.96.214204}%
  \BibitemOpen
  \bibfield  {author} {\bibinfo {author} {\bibfnamefont {M.}~\bibnamefont
  {Sonner}}, \bibinfo {author} {\bibfnamefont {K.~S.}\ \bibnamefont
  {Tikhonov}},\ and\ \bibinfo {author} {\bibfnamefont {A.~D.}\ \bibnamefont
  {Mirlin}},\ }\bibfield  {title} {\bibinfo {title} {{M}ultifractality of wave
  functions on a {C}ayley tree: {F}rom root to leaves},\ }\href
  {https://doi.org/10.1103/PhysRevB.96.214204} {\bibfield  {journal} {\bibinfo
  {journal} {Phys. Rev. B}\ }\textbf {\bibinfo {volume} {96}},\ \bibinfo
  {pages} {214204} (\bibinfo {year} {2017})}\BibitemShut {NoStop}%
\bibitem [{\citenamefont {Garc\'{\i}a-Mata}\ \emph {et~al.}(2017)\citenamefont
  {Garc\'{\i}a-Mata}, \citenamefont {Giraud}, \citenamefont {Georgeot},
  \citenamefont {Martin}, \citenamefont {Dubertrand},\ and\ \citenamefont
  {Lemari\'e}}]{PhysRevLett.118.166801}%
  \BibitemOpen
  \bibfield  {author} {\bibinfo {author} {\bibfnamefont {I.}~\bibnamefont
  {Garc\'{\i}a-Mata}}, \bibinfo {author} {\bibfnamefont {O.}~\bibnamefont
  {Giraud}}, \bibinfo {author} {\bibfnamefont {B.}~\bibnamefont {Georgeot}},
  \bibinfo {author} {\bibfnamefont {J.}~\bibnamefont {Martin}}, \bibinfo
  {author} {\bibfnamefont {R.}~\bibnamefont {Dubertrand}},\ and\ \bibinfo
  {author} {\bibfnamefont {G.}~\bibnamefont {Lemari\'e}},\ }\bibfield  {title}
  {\bibinfo {title} {{S}caling {T}heory of the {A}nderson {T}ransition in
  {R}andom {G}raphs: {E}rgodicity and {U}niversality},\ }\href
  {https://doi.org/10.1103/PhysRevLett.118.166801} {\bibfield  {journal}
  {\bibinfo  {journal} {Phys. Rev. Lett.}\ }\textbf {\bibinfo {volume} {118}},\
  \bibinfo {pages} {166801} (\bibinfo {year} {2017})}\BibitemShut {NoStop}%
\bibitem [{\citenamefont {Biroli}\ and\ \citenamefont
  {Tarzia}(2017)}]{PhysRevB.96.201114}%
  \BibitemOpen
  \bibfield  {author} {\bibinfo {author} {\bibfnamefont {G.}~\bibnamefont
  {Biroli}}\ and\ \bibinfo {author} {\bibfnamefont {M.}~\bibnamefont
  {Tarzia}},\ }\bibfield  {title} {\bibinfo {title} {{D}elocalized glassy
  dynamics and many-body localization},\ }\href
  {https://doi.org/10.1103/PhysRevB.96.201114} {\bibfield  {journal} {\bibinfo
  {journal} {Phys. Rev. B}\ }\textbf {\bibinfo {volume} {96}},\ \bibinfo
  {pages} {201114} (\bibinfo {year} {2017})}\BibitemShut {NoStop}%
\bibitem [{\citenamefont {Kravtsov}\ \emph {et~al.}(2018)\citenamefont
  {Kravtsov}, \citenamefont {Altshuler},\ and\ \citenamefont
  {Ioffe}}]{KRAVTSOV2018148}%
  \BibitemOpen
  \bibfield  {author} {\bibinfo {author} {\bibfnamefont {V.}~\bibnamefont
  {Kravtsov}}, \bibinfo {author} {\bibfnamefont {B.}~\bibnamefont
  {Altshuler}},\ and\ \bibinfo {author} {\bibfnamefont {L.}~\bibnamefont
  {Ioffe}},\ }\bibfield  {title} {\bibinfo {title} {{N}on-ergodic delocalized
  phase in {A}nderson model on {B}ethe lattice and regular graph},\ }\href
  {https://doi.org/https://doi.org/10.1016/j.aop.2017.12.009} {\bibfield
  {journal} {\bibinfo  {journal} {Annals of Physics}\ }\textbf {\bibinfo
  {volume} {389}},\ \bibinfo {pages} {148} (\bibinfo {year}
  {2018})}\BibitemShut {NoStop}%
\bibitem [{\citenamefont {Tikhonov}\ and\ \citenamefont
  {Mirlin}(2019{\natexlab{a}})}]{PhysRevB.99.214202}%
  \BibitemOpen
  \bibfield  {author} {\bibinfo {author} {\bibfnamefont {K.~S.}\ \bibnamefont
  {Tikhonov}}\ and\ \bibinfo {author} {\bibfnamefont {A.~D.}\ \bibnamefont
  {Mirlin}},\ }\bibfield  {title} {\bibinfo {title} {{C}ritical behavior at the
  localization transition on random regular graphs},\ }\href
  {https://doi.org/10.1103/PhysRevB.99.214202} {\bibfield  {journal} {\bibinfo
  {journal} {Phys. Rev. B}\ }\textbf {\bibinfo {volume} {99}},\ \bibinfo
  {pages} {214202} (\bibinfo {year} {2019}{\natexlab{a}})}\BibitemShut
  {NoStop}%
\bibitem [{\citenamefont {De~Tomasi}\ \emph {et~al.}(2020)\citenamefont
  {De~Tomasi}, \citenamefont {Bera}, \citenamefont {Scardicchio},\ and\
  \citenamefont {Khaymovich}}]{PhysRevB.101.100201}%
  \BibitemOpen
  \bibfield  {author} {\bibinfo {author} {\bibfnamefont {G.}~\bibnamefont
  {De~Tomasi}}, \bibinfo {author} {\bibfnamefont {S.}~\bibnamefont {Bera}},
  \bibinfo {author} {\bibfnamefont {A.}~\bibnamefont {Scardicchio}},\ and\
  \bibinfo {author} {\bibfnamefont {I.~M.}\ \bibnamefont {Khaymovich}},\
  }\bibfield  {title} {\bibinfo {title} {{S}ubdiffusion in the {A}nderson model
  on the random regular graph},\ }\href
  {https://doi.org/10.1103/PhysRevB.101.100201} {\bibfield  {journal} {\bibinfo
   {journal} {Phys. Rev. B}\ }\textbf {\bibinfo {volume} {101}},\ \bibinfo
  {pages} {100201} (\bibinfo {year} {2020})}\BibitemShut {NoStop}%
\bibitem [{\citenamefont {Parisi}\ \emph {et~al.}(2019)\citenamefont {Parisi},
  \citenamefont {Pascazio}, \citenamefont {Pietracaprina}, \citenamefont
  {Ros},\ and\ \citenamefont {Scardicchio}}]{Parisi_2020}%
  \BibitemOpen
  \bibfield  {author} {\bibinfo {author} {\bibfnamefont {G.}~\bibnamefont
  {Parisi}}, \bibinfo {author} {\bibfnamefont {S.}~\bibnamefont {Pascazio}},
  \bibinfo {author} {\bibfnamefont {F.}~\bibnamefont {Pietracaprina}}, \bibinfo
  {author} {\bibfnamefont {V.}~\bibnamefont {Ros}},\ and\ \bibinfo {author}
  {\bibfnamefont {A.}~\bibnamefont {Scardicchio}},\ }\bibfield  {title}
  {\bibinfo {title} {{A}nderson transition on the {B}ethe lattice: an approach
  with real energies},\ }\href {https://doi.org/10.1088/1751-8121/ab56e8}
  {\bibfield  {journal} {\bibinfo  {journal} {Journal of Physics A:
  Mathematical and Theoretical}\ }\textbf {\bibinfo {volume} {53}},\ \bibinfo
  {pages} {014003} (\bibinfo {year} {2019})}\BibitemShut {NoStop}%
\bibitem [{\citenamefont {Bera}\ \emph {et~al.}(2018)\citenamefont {Bera},
  \citenamefont {De~Tomasi}, \citenamefont {Khaymovich},\ and\ \citenamefont
  {Scardicchio}}]{PhysRevB.98.134205}%
  \BibitemOpen
  \bibfield  {author} {\bibinfo {author} {\bibfnamefont {S.}~\bibnamefont
  {Bera}}, \bibinfo {author} {\bibfnamefont {G.}~\bibnamefont {De~Tomasi}},
  \bibinfo {author} {\bibfnamefont {I.~M.}\ \bibnamefont {Khaymovich}},\ and\
  \bibinfo {author} {\bibfnamefont {A.}~\bibnamefont {Scardicchio}},\
  }\bibfield  {title} {\bibinfo {title} {{R}eturn probability for the
  {A}nderson model on the random regular graph},\ }\href
  {https://doi.org/10.1103/PhysRevB.98.134205} {\bibfield  {journal} {\bibinfo
  {journal} {Phys. Rev. B}\ }\textbf {\bibinfo {volume} {98}},\ \bibinfo
  {pages} {134205} (\bibinfo {year} {2018})}\BibitemShut {NoStop}%
\bibitem [{\citenamefont {Tikhonov}\ and\ \citenamefont
  {Mirlin}(2019{\natexlab{b}})}]{PhysRevB.99.024202}%
  \BibitemOpen
  \bibfield  {author} {\bibinfo {author} {\bibfnamefont {K.~S.}\ \bibnamefont
  {Tikhonov}}\ and\ \bibinfo {author} {\bibfnamefont {A.~D.}\ \bibnamefont
  {Mirlin}},\ }\bibfield  {title} {\bibinfo {title} {{S}tatistics of
  eigenstates near the localization transition on random regular graphs},\
  }\href {https://doi.org/10.1103/PhysRevB.99.024202} {\bibfield  {journal}
  {\bibinfo  {journal} {Phys. Rev. B}\ }\textbf {\bibinfo {volume} {99}},\
  \bibinfo {pages} {024202} (\bibinfo {year} {2019}{\natexlab{b}})}\BibitemShut
  {NoStop}%
\bibitem [{\citenamefont {Roy}\ and\ \citenamefont
  {Logan}(2020)}]{PhysRevLett.125.250402}%
  \BibitemOpen
  \bibfield  {author} {\bibinfo {author} {\bibfnamefont {S.}~\bibnamefont
  {Roy}}\ and\ \bibinfo {author} {\bibfnamefont {D.~E.}\ \bibnamefont
  {Logan}},\ }\bibfield  {title} {\bibinfo {title} {{L}ocalization on {C}ertain
  {G}raphs with {S}trongly {C}orrelated {D}isorder},\ }\href
  {https://doi.org/10.1103/PhysRevLett.125.250402} {\bibfield  {journal}
  {\bibinfo  {journal} {Phys. Rev. Lett.}\ }\textbf {\bibinfo {volume} {125}},\
  \bibinfo {pages} {250402} (\bibinfo {year} {2020})}\BibitemShut {NoStop}%
\bibitem [{\citenamefont {Biroli}\ \emph {et~al.}(2022)\citenamefont {Biroli},
  \citenamefont {Hartmann},\ and\ \citenamefont
  {Tarzia}}]{PhysRevB.105.094202}%
  \BibitemOpen
  \bibfield  {author} {\bibinfo {author} {\bibfnamefont {G.}~\bibnamefont
  {Biroli}}, \bibinfo {author} {\bibfnamefont {A.~K.}\ \bibnamefont
  {Hartmann}},\ and\ \bibinfo {author} {\bibfnamefont {M.}~\bibnamefont
  {Tarzia}},\ }\bibfield  {title} {\bibinfo {title} {{C}ritical behavior of the
  {A}nderson model on the {B}ethe lattice via a large-deviation approach},\
  }\href {https://doi.org/10.1103/PhysRevB.105.094202} {\bibfield  {journal}
  {\bibinfo  {journal} {Phys. Rev. B}\ }\textbf {\bibinfo {volume} {105}},\
  \bibinfo {pages} {094202} (\bibinfo {year} {2022})}\BibitemShut {NoStop}%
\bibitem [{\citenamefont {Tikhonov}\ and\ \citenamefont
  {Mirlin}(2021)}]{TIKHONOV2021168525}%
  \BibitemOpen
  \bibfield  {author} {\bibinfo {author} {\bibfnamefont {K.}~\bibnamefont
  {Tikhonov}}\ and\ \bibinfo {author} {\bibfnamefont {A.}~\bibnamefont
  {Mirlin}},\ }\bibfield  {title} {\bibinfo {title} {{F}rom {A}nderson
  localization on random regular graphs to many-body localization},\ }\href
  {https://doi.org/https://doi.org/10.1016/j.aop.2021.168525} {\bibfield
  {journal} {\bibinfo  {journal} {Annals of Physics}\ }\textbf {\bibinfo
  {volume} {435}},\ \bibinfo {pages} {168525} (\bibinfo {year} {2021})},\
  \bibinfo {note} {special Issue on Localisation 2020}\BibitemShut {NoStop}%
\bibitem [{\citenamefont {Garc\'{\i}a-Mata}\ \emph {et~al.}(2020)\citenamefont
  {Garc\'{\i}a-Mata}, \citenamefont {Martin}, \citenamefont {Dubertrand},
  \citenamefont {Giraud}, \citenamefont {Georgeot},\ and\ \citenamefont
  {Lemari\'e}}]{PhysRevResearch.2.012020}%
  \BibitemOpen
  \bibfield  {author} {\bibinfo {author} {\bibfnamefont {I.}~\bibnamefont
  {Garc\'{\i}a-Mata}}, \bibinfo {author} {\bibfnamefont {J.}~\bibnamefont
  {Martin}}, \bibinfo {author} {\bibfnamefont {R.}~\bibnamefont {Dubertrand}},
  \bibinfo {author} {\bibfnamefont {O.}~\bibnamefont {Giraud}}, \bibinfo
  {author} {\bibfnamefont {B.}~\bibnamefont {Georgeot}},\ and\ \bibinfo
  {author} {\bibfnamefont {G.}~\bibnamefont {Lemari\'e}},\ }\bibfield  {title}
  {\bibinfo {title} {{T}wo critical localization lengths in the {A}nderson
  transition on random graphs},\ }\href
  {https://doi.org/10.1103/PhysRevResearch.2.012020} {\bibfield  {journal}
  {\bibinfo  {journal} {Phys. Rev. Res.}\ }\textbf {\bibinfo {volume} {2}},\
  \bibinfo {pages} {012020} (\bibinfo {year} {2020})}\BibitemShut {NoStop}%
\bibitem [{\citenamefont {Colmenarez}\ \emph {et~al.}(2022)\citenamefont
  {Colmenarez}, \citenamefont {Luitz}, \citenamefont {Khaymovich},\ and\
  \citenamefont {De~Tomasi}}]{PhysRevB.105.174207}%
  \BibitemOpen
  \bibfield  {author} {\bibinfo {author} {\bibfnamefont {L.}~\bibnamefont
  {Colmenarez}}, \bibinfo {author} {\bibfnamefont {D.~J.}\ \bibnamefont
  {Luitz}}, \bibinfo {author} {\bibfnamefont {I.~M.}\ \bibnamefont
  {Khaymovich}},\ and\ \bibinfo {author} {\bibfnamefont {G.}~\bibnamefont
  {De~Tomasi}},\ }\bibfield  {title} {\bibinfo {title} {{S}ubdiffusive
  {T}houless time scaling in the {A}nderson model on random regular graphs},\
  }\href {https://doi.org/10.1103/PhysRevB.105.174207} {\bibfield  {journal}
  {\bibinfo  {journal} {Phys. Rev. B}\ }\textbf {\bibinfo {volume} {105}},\
  \bibinfo {pages} {174207} (\bibinfo {year} {2022})}\BibitemShut {NoStop}%
\bibitem [{\citenamefont {Garc\'{\i}a-Mata}\ \emph {et~al.}(2022)\citenamefont
  {Garc\'{\i}a-Mata}, \citenamefont {Martin}, \citenamefont {Giraud},
  \citenamefont {Georgeot}, \citenamefont {Dubertrand},\ and\ \citenamefont
  {Lemari\'e}}]{PhysRevB.106.214202}%
  \BibitemOpen
  \bibfield  {author} {\bibinfo {author} {\bibfnamefont {I.}~\bibnamefont
  {Garc\'{\i}a-Mata}}, \bibinfo {author} {\bibfnamefont {J.}~\bibnamefont
  {Martin}}, \bibinfo {author} {\bibfnamefont {O.}~\bibnamefont {Giraud}},
  \bibinfo {author} {\bibfnamefont {B.}~\bibnamefont {Georgeot}}, \bibinfo
  {author} {\bibfnamefont {R.}~\bibnamefont {Dubertrand}},\ and\ \bibinfo
  {author} {\bibfnamefont {G.}~\bibnamefont {Lemari\'e}},\ }\bibfield  {title}
  {\bibinfo {title} {{C}ritical properties of the {A}nderson transition on
  random graphs: {T}wo-parameter scaling theory, {K}osterlitz-{T}houless type
  flow, and many-body localization},\ }\href
  {https://doi.org/10.1103/PhysRevB.106.214202} {\bibfield  {journal} {\bibinfo
   {journal} {Phys. Rev. B}\ }\textbf {\bibinfo {volume} {106}},\ \bibinfo
  {pages} {214202} (\bibinfo {year} {2022})}\BibitemShut {NoStop}%
\bibitem [{\citenamefont {Vanoni}\ \emph {et~al.}(2023)\citenamefont {Vanoni},
  \citenamefont {Altshuler}, \citenamefont {Kravtsov},\ and\ \citenamefont
  {Scardicchio}}]{vanoni2023renormalization}%
  \BibitemOpen
  \bibfield  {author} {\bibinfo {author} {\bibfnamefont {C.}~\bibnamefont
  {Vanoni}}, \bibinfo {author} {\bibfnamefont {B.~L.}\ \bibnamefont
  {Altshuler}}, \bibinfo {author} {\bibfnamefont {V.~E.}\ \bibnamefont
  {Kravtsov}},\ and\ \bibinfo {author} {\bibfnamefont {A.}~\bibnamefont
  {Scardicchio}},\ }\href@noop {} {\bibinfo {title} {{R}enormalization {G}roup
  {A}nalysis of the {A}nderson {M}odel on {R}andom {R}egular {G}raphs}}
  (\bibinfo {year} {2023}),\ \Eprint {https://arxiv.org/abs/2306.14965}
  {arXiv:2306.14965 [cond-mat.dis-nn]} \BibitemShut {NoStop}%
\bibitem [{\citenamefont {Baroni}\ \emph {et~al.}(2023)\citenamefont {Baroni},
  \citenamefont {Lorenzana}, \citenamefont {Rizzo},\ and\ \citenamefont
  {Tarzia}}]{baroni2023corrections}%
  \BibitemOpen
  \bibfield  {author} {\bibinfo {author} {\bibfnamefont {M.}~\bibnamefont
  {Baroni}}, \bibinfo {author} {\bibfnamefont {G.~G.}\ \bibnamefont
  {Lorenzana}}, \bibinfo {author} {\bibfnamefont {T.}~\bibnamefont {Rizzo}},\
  and\ \bibinfo {author} {\bibfnamefont {M.}~\bibnamefont {Tarzia}},\
  }\href@noop {} {\bibinfo {title} {{C}orrections to the {B}ethe lattice
  solution of {A}nderson localization}} (\bibinfo {year} {2023}),\ \Eprint
  {https://arxiv.org/abs/2304.10365} {arXiv:2304.10365 [cond-mat.dis-nn]}
  \BibitemShut {NoStop}%
\bibitem [{\citenamefont {Fyodorov}\ \emph {et~al.}(1992)\citenamefont
  {Fyodorov}, \citenamefont {Mirlin},\ and\ \citenamefont
  {Sommers}}]{fyodorov1992novel}%
  \BibitemOpen
  \bibfield  {author} {\bibinfo {author} {\bibfnamefont {Y.}~\bibnamefont
  {Fyodorov}}, \bibinfo {author} {\bibfnamefont {A.}~\bibnamefont {Mirlin}},\
  and\ \bibinfo {author} {\bibfnamefont {H.-J.}\ \bibnamefont {Sommers}},\
  }\bibfield  {title} {\bibinfo {title} {{A} novel field theoretical approach
  to the {A}nderson localization: sparse random hopping model},\ }\href@noop {}
  {\bibfield  {journal} {\bibinfo  {journal} {Journal de Physique I}\ }\textbf
  {\bibinfo {volume} {2}},\ \bibinfo {pages} {1571} (\bibinfo {year}
  {1992})}\BibitemShut {NoStop}%
\bibitem [{\citenamefont {Mirlin}\ and\ \citenamefont
  {Fyodorov}(1991)}]{ADMirlin_1991}%
  \BibitemOpen
  \bibfield  {author} {\bibinfo {author} {\bibfnamefont {A.~D.}\ \bibnamefont
  {Mirlin}}\ and\ \bibinfo {author} {\bibfnamefont {Y.~V.}\ \bibnamefont
  {Fyodorov}},\ }\bibfield  {title} {\bibinfo {title} {{U}niversality of level
  correlation function of sparse random matrices},\ }\href
  {https://doi.org/10.1088/0305-4470/24/10/016} {\bibfield  {journal} {\bibinfo
   {journal} {Journal of Physics A: Mathematical and General}\ }\textbf
  {\bibinfo {volume} {24}},\ \bibinfo {pages} {2273} (\bibinfo {year}
  {1991})}\BibitemShut {NoStop}%
\bibitem [{\citenamefont {De~Roeck}\ and\ \citenamefont
  {Huveneers}(2017)}]{PhysRevB.95.155129}%
  \BibitemOpen
  \bibfield  {author} {\bibinfo {author} {\bibfnamefont {W.}~\bibnamefont
  {De~Roeck}}\ and\ \bibinfo {author} {\bibfnamefont {F.~m.~c.}\ \bibnamefont
  {Huveneers}},\ }\bibfield  {title} {\bibinfo {title} {{S}tability and
  instability towards delocalization in many-body localization systems},\
  }\href {https://doi.org/10.1103/PhysRevB.95.155129} {\bibfield  {journal}
  {\bibinfo  {journal} {Phys. Rev. B}\ }\textbf {\bibinfo {volume} {95}},\
  \bibinfo {pages} {155129} (\bibinfo {year} {2017})}\BibitemShut {NoStop}%
\bibitem [{\citenamefont {Morningstar}\ \emph {et~al.}(2022)\citenamefont
  {Morningstar}, \citenamefont {Colmenarez}, \citenamefont {Khemani},
  \citenamefont {Luitz},\ and\ \citenamefont {Huse}}]{PhysRevB.105.174205}%
  \BibitemOpen
  \bibfield  {author} {\bibinfo {author} {\bibfnamefont {A.}~\bibnamefont
  {Morningstar}}, \bibinfo {author} {\bibfnamefont {L.}~\bibnamefont
  {Colmenarez}}, \bibinfo {author} {\bibfnamefont {V.}~\bibnamefont {Khemani}},
  \bibinfo {author} {\bibfnamefont {D.~J.}\ \bibnamefont {Luitz}},\ and\
  \bibinfo {author} {\bibfnamefont {D.~A.}\ \bibnamefont {Huse}},\ }\bibfield
  {title} {\bibinfo {title} {{A}valanches and many-body resonances in many-body
  localized systems},\ }\href {https://doi.org/10.1103/PhysRevB.105.174205}
  {\bibfield  {journal} {\bibinfo  {journal} {Phys. Rev. B}\ }\textbf {\bibinfo
  {volume} {105}},\ \bibinfo {pages} {174205} (\bibinfo {year}
  {2022})}\BibitemShut {NoStop}%
\bibitem [{\citenamefont {Sels}(2022)}]{PhysRevB.106.L020202}%
  \BibitemOpen
  \bibfield  {author} {\bibinfo {author} {\bibfnamefont {D.}~\bibnamefont
  {Sels}},\ }\bibfield  {title} {\bibinfo {title} {{B}ath-induced
  delocalization in interacting disordered spin chains},\ }\href
  {https://doi.org/10.1103/PhysRevB.106.L020202} {\bibfield  {journal}
  {\bibinfo  {journal} {Phys. Rev. B}\ }\textbf {\bibinfo {volume} {106}},\
  \bibinfo {pages} {L020202} (\bibinfo {year} {2022})}\BibitemShut {NoStop}%
\bibitem [{\citenamefont {Léonard}\ \emph {et~al.}(2022)\citenamefont
  {Léonard}, \citenamefont {Kim}, \citenamefont {Rispoli}, \citenamefont
  {Lukin}, \citenamefont {Schittko}, \citenamefont {Kwan}, \citenamefont
  {Demler}, \citenamefont {Sels},\ and\ \citenamefont
  {Greiner}}]{leonard2022signatures}%
  \BibitemOpen
  \bibfield  {author} {\bibinfo {author} {\bibfnamefont {J.}~\bibnamefont
  {Léonard}}, \bibinfo {author} {\bibfnamefont {S.}~\bibnamefont {Kim}},
  \bibinfo {author} {\bibfnamefont {M.}~\bibnamefont {Rispoli}}, \bibinfo
  {author} {\bibfnamefont {A.}~\bibnamefont {Lukin}}, \bibinfo {author}
  {\bibfnamefont {R.}~\bibnamefont {Schittko}}, \bibinfo {author}
  {\bibfnamefont {J.}~\bibnamefont {Kwan}}, \bibinfo {author} {\bibfnamefont
  {E.}~\bibnamefont {Demler}}, \bibinfo {author} {\bibfnamefont
  {D.}~\bibnamefont {Sels}},\ and\ \bibinfo {author} {\bibfnamefont
  {M.}~\bibnamefont {Greiner}},\ }\href@noop {} {\bibinfo {title} {{S}ignatures
  of bath-induced quantum avalanches in a many-body--localized system}}
  (\bibinfo {year} {2022}),\ \Eprint {https://arxiv.org/abs/2012.15270}
  {arXiv:2012.15270 [cond-mat.quant-gas]} \BibitemShut {NoStop}%
\bibitem [{\citenamefont {Weiner}\ \emph {et~al.}(2019)\citenamefont {Weiner},
  \citenamefont {Evers},\ and\ \citenamefont {Bera}}]{PhysRevB.100.104204}%
  \BibitemOpen
  \bibfield  {author} {\bibinfo {author} {\bibfnamefont {F.}~\bibnamefont
  {Weiner}}, \bibinfo {author} {\bibfnamefont {F.}~\bibnamefont {Evers}},\ and\
  \bibinfo {author} {\bibfnamefont {S.}~\bibnamefont {Bera}},\ }\bibfield
  {title} {\bibinfo {title} {{S}low dynamics and strong finite-size effects in
  many-body localization with random and quasiperiodic potentials},\ }\href
  {https://doi.org/10.1103/PhysRevB.100.104204} {\bibfield  {journal} {\bibinfo
   {journal} {Phys. Rev. B}\ }\textbf {\bibinfo {volume} {100}},\ \bibinfo
  {pages} {104204} (\bibinfo {year} {2019})}\BibitemShut {NoStop}%
\bibitem [{\citenamefont {\ifmmode~\check{S}\else \v{S}\fi{}untajs}\ \emph
  {et~al.}(2020{\natexlab{a}})\citenamefont {\ifmmode~\check{S}\else
  \v{S}\fi{}untajs}, \citenamefont {Bon\ifmmode~\check{c}\else \v{c}\fi{}a},
  \citenamefont {Prosen},\ and\ \citenamefont {Vidmar}}]{PhysRevE.102.062144}%
  \BibitemOpen
  \bibfield  {author} {\bibinfo {author} {\bibfnamefont {J.}~\bibnamefont
  {\ifmmode~\check{S}\else \v{S}\fi{}untajs}}, \bibinfo {author} {\bibfnamefont
  {J.}~\bibnamefont {Bon\ifmmode~\check{c}\else \v{c}\fi{}a}}, \bibinfo
  {author} {\bibfnamefont {T.~c.~v.}\ \bibnamefont {Prosen}},\ and\ \bibinfo
  {author} {\bibfnamefont {L.}~\bibnamefont {Vidmar}},\ }\bibfield  {title}
  {\bibinfo {title} {{Q}uantum chaos challenges many-body localization},\
  }\href {https://doi.org/10.1103/PhysRevE.102.062144} {\bibfield  {journal}
  {\bibinfo  {journal} {Phys. Rev. E}\ }\textbf {\bibinfo {volume} {102}},\
  \bibinfo {pages} {062144} (\bibinfo {year} {2020}{\natexlab{a}})}\BibitemShut
  {NoStop}%
\bibitem [{\citenamefont {\ifmmode~\check{S}\else \v{S}\fi{}untajs}\ \emph
  {et~al.}(2020{\natexlab{b}})\citenamefont {\ifmmode~\check{S}\else
  \v{S}\fi{}untajs}, \citenamefont {Bon\ifmmode~\check{c}\else \v{c}\fi{}a},
  \citenamefont {Prosen},\ and\ \citenamefont {Vidmar}}]{PhysRevB.102.064207}%
  \BibitemOpen
  \bibfield  {author} {\bibinfo {author} {\bibfnamefont {J.}~\bibnamefont
  {\ifmmode~\check{S}\else \v{S}\fi{}untajs}}, \bibinfo {author} {\bibfnamefont
  {J.}~\bibnamefont {Bon\ifmmode~\check{c}\else \v{c}\fi{}a}}, \bibinfo
  {author} {\bibfnamefont {T.~c.~v.}\ \bibnamefont {Prosen}},\ and\ \bibinfo
  {author} {\bibfnamefont {L.}~\bibnamefont {Vidmar}},\ }\bibfield  {title}
  {\bibinfo {title} {{E}rgodicity breaking transition in finite disordered spin
  chains},\ }\href {https://doi.org/10.1103/PhysRevB.102.064207} {\bibfield
  {journal} {\bibinfo  {journal} {Phys. Rev. B}\ }\textbf {\bibinfo {volume}
  {102}},\ \bibinfo {pages} {064207} (\bibinfo {year}
  {2020}{\natexlab{b}})}\BibitemShut {NoStop}%
\bibitem [{\citenamefont {Kiefer-Emmanouilidis}\ \emph
  {et~al.}(2021)\citenamefont {Kiefer-Emmanouilidis}, \citenamefont {Unanyan},
  \citenamefont {Fleischhauer},\ and\ \citenamefont
  {Sirker}}]{PhysRevB.103.024203}%
  \BibitemOpen
  \bibfield  {author} {\bibinfo {author} {\bibfnamefont {M.}~\bibnamefont
  {Kiefer-Emmanouilidis}}, \bibinfo {author} {\bibfnamefont {R.}~\bibnamefont
  {Unanyan}}, \bibinfo {author} {\bibfnamefont {M.}~\bibnamefont
  {Fleischhauer}},\ and\ \bibinfo {author} {\bibfnamefont {J.}~\bibnamefont
  {Sirker}},\ }\bibfield  {title} {\bibinfo {title} {{S}low delocalization of
  particles in many-body localized phases},\ }\href
  {https://doi.org/10.1103/PhysRevB.103.024203} {\bibfield  {journal} {\bibinfo
   {journal} {Phys. Rev. B}\ }\textbf {\bibinfo {volume} {103}},\ \bibinfo
  {pages} {024203} (\bibinfo {year} {2021})}\BibitemShut {NoStop}%
\bibitem [{\citenamefont {Sels}\ and\ \citenamefont
  {Polkovnikov}(2021)}]{PhysRevE.104.054105}%
  \BibitemOpen
  \bibfield  {author} {\bibinfo {author} {\bibfnamefont {D.}~\bibnamefont
  {Sels}}\ and\ \bibinfo {author} {\bibfnamefont {A.}~\bibnamefont
  {Polkovnikov}},\ }\bibfield  {title} {\bibinfo {title} {{D}ynamical
  obstruction to localization in a disordered spin chain},\ }\href
  {https://doi.org/10.1103/PhysRevE.104.054105} {\bibfield  {journal} {\bibinfo
   {journal} {Phys. Rev. E}\ }\textbf {\bibinfo {volume} {104}},\ \bibinfo
  {pages} {054105} (\bibinfo {year} {2021})}\BibitemShut {NoStop}%
\bibitem [{\citenamefont {Vidmar}\ \emph {et~al.}(2021)\citenamefont {Vidmar},
  \citenamefont {Krajewski}, \citenamefont {Bon\ifmmode~\check{c}\else
  \v{c}\fi{}a},\ and\ \citenamefont {Mierzejewski}}]{PhysRevLett.127.230603}%
  \BibitemOpen
  \bibfield  {author} {\bibinfo {author} {\bibfnamefont {L.}~\bibnamefont
  {Vidmar}}, \bibinfo {author} {\bibfnamefont {B.}~\bibnamefont {Krajewski}},
  \bibinfo {author} {\bibfnamefont {J.}~\bibnamefont
  {Bon\ifmmode~\check{c}\else \v{c}\fi{}a}},\ and\ \bibinfo {author}
  {\bibfnamefont {M.}~\bibnamefont {Mierzejewski}},\ }\bibfield  {title}
  {\bibinfo {title} {{P}henomenology of {S}pectral {F}unctions in {D}isordered
  {S}pin {C}hains at {I}nfinite {T}emperature},\ }\href
  {https://doi.org/10.1103/PhysRevLett.127.230603} {\bibfield  {journal}
  {\bibinfo  {journal} {Phys. Rev. Lett.}\ }\textbf {\bibinfo {volume} {127}},\
  \bibinfo {pages} {230603} (\bibinfo {year} {2021})}\BibitemShut {NoStop}%
\bibitem [{\citenamefont {Abanin}\ \emph {et~al.}(2021)\citenamefont {Abanin},
  \citenamefont {Bardarson}, \citenamefont {{De Tomasi}}, \citenamefont
  {Gopalakrishnan}, \citenamefont {Khemani}, \citenamefont {Parameswaran},
  \citenamefont {Pollmann}, \citenamefont {Potter}, \citenamefont {Serbyn},\
  and\ \citenamefont {Vasseur}}]{ABANIN2021168415}%
  \BibitemOpen
  \bibfield  {author} {\bibinfo {author} {\bibfnamefont {D.}~\bibnamefont
  {Abanin}}, \bibinfo {author} {\bibfnamefont {J.}~\bibnamefont {Bardarson}},
  \bibinfo {author} {\bibfnamefont {G.}~\bibnamefont {{De Tomasi}}}, \bibinfo
  {author} {\bibfnamefont {S.}~\bibnamefont {Gopalakrishnan}}, \bibinfo
  {author} {\bibfnamefont {V.}~\bibnamefont {Khemani}}, \bibinfo {author}
  {\bibfnamefont {S.}~\bibnamefont {Parameswaran}}, \bibinfo {author}
  {\bibfnamefont {F.}~\bibnamefont {Pollmann}}, \bibinfo {author}
  {\bibfnamefont {A.}~\bibnamefont {Potter}}, \bibinfo {author} {\bibfnamefont
  {M.}~\bibnamefont {Serbyn}},\ and\ \bibinfo {author} {\bibfnamefont
  {R.}~\bibnamefont {Vasseur}},\ }\bibfield  {title} {\bibinfo {title}
  {{D}istinguishing localization from chaos: {C}hallenges in finite-size
  systems},\ }\href {https://doi.org/https://doi.org/10.1016/j.aop.2021.168415}
  {\bibfield  {journal} {\bibinfo  {journal} {Annals of Physics}\ }\textbf
  {\bibinfo {volume} {427}},\ \bibinfo {pages} {168415} (\bibinfo {year}
  {2021})}\BibitemShut {NoStop}%
\bibitem [{\citenamefont {Sierant}\ \emph {et~al.}(2020)\citenamefont
  {Sierant}, \citenamefont {Delande},\ and\ \citenamefont
  {Zakrzewski}}]{PhysRevLett.124.186601}%
  \BibitemOpen
  \bibfield  {author} {\bibinfo {author} {\bibfnamefont {P.}~\bibnamefont
  {Sierant}}, \bibinfo {author} {\bibfnamefont {D.}~\bibnamefont {Delande}},\
  and\ \bibinfo {author} {\bibfnamefont {J.}~\bibnamefont {Zakrzewski}},\
  }\bibfield  {title} {\bibinfo {title} {{T}houless {T}ime {A}nalysis of
  {A}nderson and {M}any-{B}ody {L}ocalization {T}ransitions},\ }\href
  {https://doi.org/10.1103/PhysRevLett.124.186601} {\bibfield  {journal}
  {\bibinfo  {journal} {Phys. Rev. Lett.}\ }\textbf {\bibinfo {volume} {124}},\
  \bibinfo {pages} {186601} (\bibinfo {year} {2020})}\BibitemShut {NoStop}%
\bibitem [{\citenamefont {Panda}\ \emph {et~al.}(2020)\citenamefont {Panda},
  \citenamefont {Scardicchio}, \citenamefont {Schulz}, \citenamefont {Taylor},\
  and\ \citenamefont {Žnidarič}}]{Panda_2019}%
  \BibitemOpen
  \bibfield  {author} {\bibinfo {author} {\bibfnamefont {R.~K.}\ \bibnamefont
  {Panda}}, \bibinfo {author} {\bibfnamefont {A.}~\bibnamefont {Scardicchio}},
  \bibinfo {author} {\bibfnamefont {M.}~\bibnamefont {Schulz}}, \bibinfo
  {author} {\bibfnamefont {S.~R.}\ \bibnamefont {Taylor}},\ and\ \bibinfo
  {author} {\bibfnamefont {M.}~\bibnamefont {Žnidarič}},\ }\bibfield  {title}
  {\bibinfo {title} {{C}an we study the many-body localisation transition?},\
  }\href {https://doi.org/10.1209/0295-5075/128/67003} {\bibfield  {journal}
  {\bibinfo  {journal} {Europhysics Letters}\ }\textbf {\bibinfo {volume}
  {128}},\ \bibinfo {pages} {67003} (\bibinfo {year} {2020})}\BibitemShut
  {NoStop}%
\bibitem [{\citenamefont {Luitz}\ and\ \citenamefont
  {Lev}(2020)}]{PhysRevB.102.100202}%
  \BibitemOpen
  \bibfield  {author} {\bibinfo {author} {\bibfnamefont {D.~J.}\ \bibnamefont
  {Luitz}}\ and\ \bibinfo {author} {\bibfnamefont {Y.~B.}\ \bibnamefont
  {Lev}},\ }\bibfield  {title} {\bibinfo {title} {{A}bsence of slow particle
  transport in the many-body localized phase},\ }\href
  {https://doi.org/10.1103/PhysRevB.102.100202} {\bibfield  {journal} {\bibinfo
   {journal} {Phys. Rev. B}\ }\textbf {\bibinfo {volume} {102}},\ \bibinfo
  {pages} {100202} (\bibinfo {year} {2020})}\BibitemShut {NoStop}%
\bibitem [{\citenamefont {Mirlin}\ \emph {et~al.}(2006)\citenamefont {Mirlin},
  \citenamefont {Fyodorov}, \citenamefont {Mildenberger},\ and\ \citenamefont
  {Evers}}]{PhysRevLett.97.046803}%
  \BibitemOpen
  \bibfield  {author} {\bibinfo {author} {\bibfnamefont {A.~D.}\ \bibnamefont
  {Mirlin}}, \bibinfo {author} {\bibfnamefont {Y.~V.}\ \bibnamefont
  {Fyodorov}}, \bibinfo {author} {\bibfnamefont {A.}~\bibnamefont
  {Mildenberger}},\ and\ \bibinfo {author} {\bibfnamefont {F.}~\bibnamefont
  {Evers}},\ }\bibfield  {title} {\bibinfo {title} {{E}xact {R}elations between
  {M}ultifractal {E}xponents at the {A}nderson {T}ransition},\ }\href
  {https://doi.org/10.1103/PhysRevLett.97.046803} {\bibfield  {journal}
  {\bibinfo  {journal} {Phys. Rev. Lett.}\ }\textbf {\bibinfo {volume} {97}},\
  \bibinfo {pages} {046803} (\bibinfo {year} {2006})}\BibitemShut {NoStop}%
\bibitem [{\citenamefont {Gruzberg}\ \emph {et~al.}(2011)\citenamefont
  {Gruzberg}, \citenamefont {Ludwig}, \citenamefont {Mirlin},\ and\
  \citenamefont {Zirnbauer}}]{gruzberg2011symmetries}%
  \BibitemOpen
  \bibfield  {author} {\bibinfo {author} {\bibfnamefont {I.~A.}\ \bibnamefont
  {Gruzberg}}, \bibinfo {author} {\bibfnamefont {A.~W.~W.}\ \bibnamefont
  {Ludwig}}, \bibinfo {author} {\bibfnamefont {A.~D.}\ \bibnamefont {Mirlin}},\
  and\ \bibinfo {author} {\bibfnamefont {M.~R.}\ \bibnamefont {Zirnbauer}},\
  }\bibfield  {title} {\bibinfo {title} {{S}ymmetries of multifractal spectra
  and field theories of {A}nderson localization},\ }\href
  {https://doi.org/10.1103/PhysRevLett.107.086403} {\bibfield  {journal}
  {\bibinfo  {journal} {Phys. Rev. Lett.}\ }\textbf {\bibinfo {volume} {107}},\
  \bibinfo {pages} {086403} (\bibinfo {year} {2011})}\BibitemShut {NoStop}%
\bibitem [{\citenamefont {Gruzberg}\ \emph {et~al.}(2013)\citenamefont
  {Gruzberg}, \citenamefont {Mirlin},\ and\ \citenamefont
  {Zirnbauer}}]{gruzberg2013classification}%
  \BibitemOpen
  \bibfield  {author} {\bibinfo {author} {\bibfnamefont {I.~A.}\ \bibnamefont
  {Gruzberg}}, \bibinfo {author} {\bibfnamefont {A.~D.}\ \bibnamefont
  {Mirlin}},\ and\ \bibinfo {author} {\bibfnamefont {M.~R.}\ \bibnamefont
  {Zirnbauer}},\ }\bibfield  {title} {\bibinfo {title} {{C}lassification and
  symmetry properties of scaling dimensions at {A}nderson transitions},\ }\href
  {https://doi.org/10.1103/PhysRevB.87.125144} {\bibfield  {journal} {\bibinfo
  {journal} {Phys. Rev. B}\ }\textbf {\bibinfo {volume} {87}},\ \bibinfo
  {pages} {125144} (\bibinfo {year} {2013})}\BibitemShut {NoStop}%
\bibitem [{\citenamefont {Bilen}\ \emph {et~al.}(2021)\citenamefont {Bilen},
  \citenamefont {Georgeot}, \citenamefont {Giraud}, \citenamefont {Lemari\'e},\
  and\ \citenamefont {Garc\'{\i}a-Mata}}]{PhysRevResearch.3.L022023}%
  \BibitemOpen
  \bibfield  {author} {\bibinfo {author} {\bibfnamefont {A.~M.}\ \bibnamefont
  {Bilen}}, \bibinfo {author} {\bibfnamefont {B.}~\bibnamefont {Georgeot}},
  \bibinfo {author} {\bibfnamefont {O.}~\bibnamefont {Giraud}}, \bibinfo
  {author} {\bibfnamefont {G.}~\bibnamefont {Lemari\'e}},\ and\ \bibinfo
  {author} {\bibfnamefont {I.}~\bibnamefont {Garc\'{\i}a-Mata}},\ }\bibfield
  {title} {\bibinfo {title} {{S}ymmetry violation of quantum multifractality:
  {G}aussian fluctuations versus algebraic localization},\ }\href
  {https://doi.org/10.1103/PhysRevResearch.3.L022023} {\bibfield  {journal}
  {\bibinfo  {journal} {Phys. Rev. Res.}\ }\textbf {\bibinfo {volume} {3}},\
  \bibinfo {pages} {L022023} (\bibinfo {year} {2021})}\BibitemShut {NoStop}%
\bibitem [{\citenamefont {Chen}\ \emph
  {et~al.}(2023{\natexlab{b}})\citenamefont {Chen}, \citenamefont {Maciejko},\
  and\ \citenamefont {Boettcher}}]{chen2023anderson}%
  \BibitemOpen
  \bibfield  {author} {\bibinfo {author} {\bibfnamefont {A.}~\bibnamefont
  {Chen}}, \bibinfo {author} {\bibfnamefont {J.}~\bibnamefont {Maciejko}},\
  and\ \bibinfo {author} {\bibfnamefont {I.}~\bibnamefont {Boettcher}},\
  }\href@noop {} {\bibinfo {title} {Anderson localization transition in
  disordered hyperbolic lattices}} (\bibinfo {year} {2023}{\natexlab{b}}),\
  \Eprint {https://arxiv.org/abs/2310.07978} {arXiv:2310.07978
  [cond-mat.dis-nn]} \BibitemShut {NoStop}%
\bibitem [{\citenamefont {Mirlin}\ \emph {et~al.}(1996)\citenamefont {Mirlin},
  \citenamefont {Fyodorov}, \citenamefont {Dittes}, \citenamefont {Quezada},\
  and\ \citenamefont {Seligman}}]{PhysRevE.54.3221}%
  \BibitemOpen
  \bibfield  {author} {\bibinfo {author} {\bibfnamefont {A.~D.}\ \bibnamefont
  {Mirlin}}, \bibinfo {author} {\bibfnamefont {Y.~V.}\ \bibnamefont
  {Fyodorov}}, \bibinfo {author} {\bibfnamefont {F.-M.}\ \bibnamefont
  {Dittes}}, \bibinfo {author} {\bibfnamefont {J.}~\bibnamefont {Quezada}},\
  and\ \bibinfo {author} {\bibfnamefont {T.~H.}\ \bibnamefont {Seligman}},\
  }\bibfield  {title} {\bibinfo {title} {{T}ransition from localized to
  extended eigenstates in the ensemble of power-law random banded matrices},\
  }\href {https://doi.org/10.1103/PhysRevE.54.3221} {\bibfield  {journal}
  {\bibinfo  {journal} {Phys. Rev. E}\ }\textbf {\bibinfo {volume} {54}},\
  \bibinfo {pages} {3221} (\bibinfo {year} {1996})}\BibitemShut {NoStop}%
\bibitem [{\citenamefont {Mirlin}\ and\ \citenamefont
  {Evers}(2000)}]{PhysRevB.62.7920}%
  \BibitemOpen
  \bibfield  {author} {\bibinfo {author} {\bibfnamefont {A.~D.}\ \bibnamefont
  {Mirlin}}\ and\ \bibinfo {author} {\bibfnamefont {F.}~\bibnamefont {Evers}},\
  }\bibfield  {title} {\bibinfo {title} {{M}ultifractality and critical
  fluctuations at the {A}nderson transition},\ }\href
  {https://doi.org/10.1103/PhysRevB.62.7920} {\bibfield  {journal} {\bibinfo
  {journal} {Phys. Rev. B}\ }\textbf {\bibinfo {volume} {62}},\ \bibinfo
  {pages} {7920} (\bibinfo {year} {2000})}\BibitemShut {NoStop}%
\bibitem [{\citenamefont {Bogomolny}\ and\ \citenamefont
  {Sieber}(2018{\natexlab{a}})}]{PhysRevE.98.042116}%
  \BibitemOpen
  \bibfield  {author} {\bibinfo {author} {\bibfnamefont {E.}~\bibnamefont
  {Bogomolny}}\ and\ \bibinfo {author} {\bibfnamefont {M.}~\bibnamefont
  {Sieber}},\ }\bibfield  {title} {\bibinfo {title} {{P}ower-law random banded
  matrices and ultrametric matrices: {E}igenvector distribution in the
  intermediate regime},\ }\href {https://doi.org/10.1103/PhysRevE.98.042116}
  {\bibfield  {journal} {\bibinfo  {journal} {Phys. Rev. E}\ }\textbf {\bibinfo
  {volume} {98}},\ \bibinfo {pages} {042116} (\bibinfo {year}
  {2018}{\natexlab{a}})}\BibitemShut {NoStop}%
\bibitem [{\citenamefont {Bogomolny}\ \emph {et~al.}(2009)\citenamefont
  {Bogomolny}, \citenamefont {Giraud},\ and\ \citenamefont
  {Schmit}}]{PhysRevLett.103.054103}%
  \BibitemOpen
  \bibfield  {author} {\bibinfo {author} {\bibfnamefont {E.}~\bibnamefont
  {Bogomolny}}, \bibinfo {author} {\bibfnamefont {O.}~\bibnamefont {Giraud}},\
  and\ \bibinfo {author} {\bibfnamefont {C.}~\bibnamefont {Schmit}},\
  }\bibfield  {title} {\bibinfo {title} {{R}andom {M}atrix {E}nsembles
  {A}ssociated with {L}ax {M}atrices},\ }\href
  {https://doi.org/10.1103/PhysRevLett.103.054103} {\bibfield  {journal}
  {\bibinfo  {journal} {Phys. Rev. Lett.}\ }\textbf {\bibinfo {volume} {103}},\
  \bibinfo {pages} {054103} (\bibinfo {year} {2009})}\BibitemShut {NoStop}%
\bibitem [{\citenamefont {Bogomolny}\ and\ \citenamefont
  {Giraud}(2011{\natexlab{a}})}]{PhysRevE.84.036212}%
  \BibitemOpen
  \bibfield  {author} {\bibinfo {author} {\bibfnamefont {E.}~\bibnamefont
  {Bogomolny}}\ and\ \bibinfo {author} {\bibfnamefont {O.}~\bibnamefont
  {Giraud}},\ }\bibfield  {title} {\bibinfo {title} {{P}erturbation approach to
  multifractal dimensions for certain critical random-matrix ensembles},\
  }\href {https://doi.org/10.1103/PhysRevE.84.036212} {\bibfield  {journal}
  {\bibinfo  {journal} {Phys. Rev. E}\ }\textbf {\bibinfo {volume} {84}},\
  \bibinfo {pages} {036212} (\bibinfo {year} {2011}{\natexlab{a}})}\BibitemShut
  {NoStop}%
\bibitem [{\citenamefont {Bogomolny}\ and\ \citenamefont
  {Giraud}(2012)}]{PhysRevE.85.046208}%
  \BibitemOpen
  \bibfield  {author} {\bibinfo {author} {\bibfnamefont {E.}~\bibnamefont
  {Bogomolny}}\ and\ \bibinfo {author} {\bibfnamefont {O.}~\bibnamefont
  {Giraud}},\ }\bibfield  {title} {\bibinfo {title} {{M}ultifractal dimensions
  for all moments for certain critical random-matrix ensembles in the strong
  multifractality regime},\ }\href {https://doi.org/10.1103/PhysRevE.85.046208}
  {\bibfield  {journal} {\bibinfo  {journal} {Phys. Rev. E}\ }\textbf {\bibinfo
  {volume} {85}},\ \bibinfo {pages} {046208} (\bibinfo {year}
  {2012})}\BibitemShut {NoStop}%
\bibitem [{\citenamefont {Levitov}(1989)}]{L.S.Levitov_1989}%
  \BibitemOpen
  \bibfield  {author} {\bibinfo {author} {\bibfnamefont {L.~S.}\ \bibnamefont
  {Levitov}},\ }\bibfield  {title} {\bibinfo {title} {{A}bsence of
  {L}ocalization of {V}ibrational {M}odes {D}ue to {D}ipole-{D}ipole
  {I}nteraction},\ }\href {https://doi.org/10.1209/0295-5075/9/1/015}
  {\bibfield  {journal} {\bibinfo  {journal} {Europhysics Letters}\ }\textbf
  {\bibinfo {volume} {9}},\ \bibinfo {pages} {83} (\bibinfo {year}
  {1989})}\BibitemShut {NoStop}%
\bibitem [{\citenamefont {Levitov}(1990)}]{PhysRevLett.64.547}%
  \BibitemOpen
  \bibfield  {author} {\bibinfo {author} {\bibfnamefont {L.~S.}\ \bibnamefont
  {Levitov}},\ }\bibfield  {title} {\bibinfo {title} {{D}elocalization of
  vibrational modes caused by electric dipole interaction},\ }\href
  {https://doi.org/10.1103/PhysRevLett.64.547} {\bibfield  {journal} {\bibinfo
  {journal} {Phys. Rev. Lett.}\ }\textbf {\bibinfo {volume} {64}},\ \bibinfo
  {pages} {547} (\bibinfo {year} {1990})}\BibitemShut {NoStop}%
\bibitem [{\citenamefont {Monthus}\ and\ \citenamefont
  {Garel}(2010)}]{Monthus_2010}%
  \BibitemOpen
  \bibfield  {author} {\bibinfo {author} {\bibfnamefont {C.}~\bibnamefont
  {Monthus}}\ and\ \bibinfo {author} {\bibfnamefont {T.}~\bibnamefont
  {Garel}},\ }\bibfield  {title} {\bibinfo {title} {{T}he {A}nderson
  localization transition with long-ranged hoppings: analysis of the strong
  multifractality regime in terms of weighted {L}évy sums},\ }\href
  {https://doi.org/10.1088/1742-5468/2010/09/P09015} {\bibfield  {journal}
  {\bibinfo  {journal} {Journal of Statistical Mechanics: Theory and
  Experiment}\ }\textbf {\bibinfo {volume} {2010}},\ \bibinfo {pages} {P09015}
  (\bibinfo {year} {2010})}\BibitemShut {NoStop}%
\bibitem [{\citenamefont {Bogomolny}\ and\ \citenamefont
  {Giraud}(2011{\natexlab{b}})}]{PhysRevLett.106.044101}%
  \BibitemOpen
  \bibfield  {author} {\bibinfo {author} {\bibfnamefont {E.}~\bibnamefont
  {Bogomolny}}\ and\ \bibinfo {author} {\bibfnamefont {O.}~\bibnamefont
  {Giraud}},\ }\bibfield  {title} {\bibinfo {title} {{E}igenfunction {E}ntropy
  and {S}pectral {C}ompressibility for {C}ritical {R}andom {M}atrix
  {E}nsembles},\ }\href {https://doi.org/10.1103/PhysRevLett.106.044101}
  {\bibfield  {journal} {\bibinfo  {journal} {Phys. Rev. Lett.}\ }\textbf
  {\bibinfo {volume} {106}},\ \bibinfo {pages} {044101} (\bibinfo {year}
  {2011}{\natexlab{b}})}\BibitemShut {NoStop}%
\bibitem [{\citenamefont {Yevtushenko}\ and\ \citenamefont
  {Ossipov}(2007)}]{Yevtushenko_2007}%
  \BibitemOpen
  \bibfield  {author} {\bibinfo {author} {\bibfnamefont {O.}~\bibnamefont
  {Yevtushenko}}\ and\ \bibinfo {author} {\bibfnamefont {A.}~\bibnamefont
  {Ossipov}},\ }\bibfield  {title} {\bibinfo {title} {{A} supersymmetry
  approach to almost diagonal random matrices},\ }\href
  {https://doi.org/10.1088/1751-8113/40/18/002} {\bibfield  {journal} {\bibinfo
   {journal} {Journal of Physics A: Mathematical and Theoretical}\ }\textbf
  {\bibinfo {volume} {40}},\ \bibinfo {pages} {4691} (\bibinfo {year}
  {2007})}\BibitemShut {NoStop}%
\bibitem [{\citenamefont {Kravtsov}\ \emph {et~al.}(2015)\citenamefont
  {Kravtsov}, \citenamefont {Khaymovich}, \citenamefont {Cuevas},\ and\
  \citenamefont {Amini}}]{Kravtsov_2015}%
  \BibitemOpen
  \bibfield  {author} {\bibinfo {author} {\bibfnamefont {V.~E.}\ \bibnamefont
  {Kravtsov}}, \bibinfo {author} {\bibfnamefont {I.~M.}\ \bibnamefont
  {Khaymovich}}, \bibinfo {author} {\bibfnamefont {E.}~\bibnamefont {Cuevas}},\
  and\ \bibinfo {author} {\bibfnamefont {M.}~\bibnamefont {Amini}},\ }\bibfield
   {title} {\bibinfo {title} {{A} random matrix model with localization and
  ergodic transitions},\ }\href
  {https://doi.org/10.1088/1367-2630/17/12/122002} {\bibfield  {journal}
  {\bibinfo  {journal} {New Journal of Physics}\ }\textbf {\bibinfo {volume}
  {17}},\ \bibinfo {pages} {122002} (\bibinfo {year} {2015})}\BibitemShut
  {NoStop}%
\bibitem [{\citenamefont {Tomasi}\ \emph {et~al.}(2019)\citenamefont {Tomasi},
  \citenamefont {Amini}, \citenamefont {Bera}, \citenamefont {Khaymovich},\
  and\ \citenamefont {Kravtsov}}]{10.21468/SciPostPhys.6.1.014}%
  \BibitemOpen
  \bibfield  {author} {\bibinfo {author} {\bibfnamefont {G.~D.}\ \bibnamefont
  {Tomasi}}, \bibinfo {author} {\bibfnamefont {M.}~\bibnamefont {Amini}},
  \bibinfo {author} {\bibfnamefont {S.}~\bibnamefont {Bera}}, \bibinfo {author}
  {\bibfnamefont {I.~M.}\ \bibnamefont {Khaymovich}},\ and\ \bibinfo {author}
  {\bibfnamefont {V.~E.}\ \bibnamefont {Kravtsov}},\ }\bibfield  {title}
  {\bibinfo {title} {{{S}urvival probability in {G}eneralized
  {R}osenzweig-{P}orter random matrix ensemble}},\ }\href
  {https://doi.org/10.21468/SciPostPhys.6.1.014} {\bibfield  {journal}
  {\bibinfo  {journal} {SciPost Phys.}\ }\textbf {\bibinfo {volume} {6}},\
  \bibinfo {pages} {014} (\bibinfo {year} {2019})}\BibitemShut {NoStop}%
\bibitem [{\citenamefont {Khaymovich}\ \emph {et~al.}(2020)\citenamefont
  {Khaymovich}, \citenamefont {Kravtsov}, \citenamefont {Altshuler},\ and\
  \citenamefont {Ioffe}}]{PhysRevResearch.2.043346}%
  \BibitemOpen
  \bibfield  {author} {\bibinfo {author} {\bibfnamefont {I.~M.}\ \bibnamefont
  {Khaymovich}}, \bibinfo {author} {\bibfnamefont {V.~E.}\ \bibnamefont
  {Kravtsov}}, \bibinfo {author} {\bibfnamefont {B.~L.}\ \bibnamefont
  {Altshuler}},\ and\ \bibinfo {author} {\bibfnamefont {L.~B.}\ \bibnamefont
  {Ioffe}},\ }\bibfield  {title} {\bibinfo {title} {{F}ragile extended phases
  in the log-normal {R}osenzweig-{P}orter model},\ }\href
  {https://doi.org/10.1103/PhysRevResearch.2.043346} {\bibfield  {journal}
  {\bibinfo  {journal} {Phys. Rev. Res.}\ }\textbf {\bibinfo {volume} {2}},\
  \bibinfo {pages} {043346} (\bibinfo {year} {2020})}\BibitemShut {NoStop}%
\bibitem [{\citenamefont {von Soosten}\ and\ \citenamefont
  {Warzel}(2019)}]{von2019non}%
  \BibitemOpen
  \bibfield  {author} {\bibinfo {author} {\bibfnamefont {P.}~\bibnamefont {von
  Soosten}}\ and\ \bibinfo {author} {\bibfnamefont {S.}~\bibnamefont
  {Warzel}},\ }\bibfield  {title} {\bibinfo {title} {{N}on-ergodic
  delocalization in the {R}osenzweig--{P}orter model},\ }\href@noop {}
  {\bibfield  {journal} {\bibinfo  {journal} {Letters in Mathematical Physics}\
  }\textbf {\bibinfo {volume} {109}},\ \bibinfo {pages} {905} (\bibinfo {year}
  {2019})}\BibitemShut {NoStop}%
\bibitem [{\citenamefont {Truong}\ and\ \citenamefont
  {Ossipov}(2016)}]{Truong_2016}%
  \BibitemOpen
  \bibfield  {author} {\bibinfo {author} {\bibfnamefont {K.}~\bibnamefont
  {Truong}}\ and\ \bibinfo {author} {\bibfnamefont {A.}~\bibnamefont
  {Ossipov}},\ }\bibfield  {title} {\bibinfo {title} {{E}igenvectors under a
  generic perturbation: {N}on-perturbative results from the random matrix
  approach},\ }\href {https://doi.org/10.1209/0295-5075/116/37002} {\bibfield
  {journal} {\bibinfo  {journal} {Europhysics Letters}\ }\textbf {\bibinfo
  {volume} {116}},\ \bibinfo {pages} {37002} (\bibinfo {year}
  {2016})}\BibitemShut {NoStop}%
\bibitem [{\citenamefont {Bogomolny}\ and\ \citenamefont
  {Sieber}(2018{\natexlab{b}})}]{PhysRevE.98.032139}%
  \BibitemOpen
  \bibfield  {author} {\bibinfo {author} {\bibfnamefont {E.}~\bibnamefont
  {Bogomolny}}\ and\ \bibinfo {author} {\bibfnamefont {M.}~\bibnamefont
  {Sieber}},\ }\bibfield  {title} {\bibinfo {title} {{E}igenfunction
  distribution for the {R}osenzweig-{P}orter model},\ }\href
  {https://doi.org/10.1103/PhysRevE.98.032139} {\bibfield  {journal} {\bibinfo
  {journal} {Phys. Rev. E}\ }\textbf {\bibinfo {volume} {98}},\ \bibinfo
  {pages} {032139} (\bibinfo {year} {2018}{\natexlab{b}})}\BibitemShut
  {NoStop}%
\bibitem [{\citenamefont {Monthus}(2017)}]{Monthus_2017}%
  \BibitemOpen
  \bibfield  {author} {\bibinfo {author} {\bibfnamefont {C.}~\bibnamefont
  {Monthus}},\ }\bibfield  {title} {\bibinfo {title} {{M}ultifractality of
  eigenstates in the delocalized non-ergodic phase of some random matrix
  models: {W}igner–{W}eisskopf approach},\ }\href
  {https://doi.org/10.1088/1751-8121/aa77e1} {\bibfield  {journal} {\bibinfo
  {journal} {Journal of Physics A: Mathematical and Theoretical}\ }\textbf
  {\bibinfo {volume} {50}},\ \bibinfo {pages} {295101} (\bibinfo {year}
  {2017})}\BibitemShut {NoStop}%
\bibitem [{\citenamefont {Amini}(2017)}]{Amini_2017}%
  \BibitemOpen
  \bibfield  {author} {\bibinfo {author} {\bibfnamefont {M.}~\bibnamefont
  {Amini}},\ }\bibfield  {title} {\bibinfo {title} {{S}pread of wave packets in
  disordered hierarchical lattices},\ }\href
  {https://doi.org/10.1209/0295-5075/117/30003} {\bibfield  {journal} {\bibinfo
   {journal} {Europhysics Letters}\ }\textbf {\bibinfo {volume} {117}},\
  \bibinfo {pages} {30003} (\bibinfo {year} {2017})}\BibitemShut {NoStop}%
\bibitem [{\citenamefont {Biroli}\ and\ \citenamefont
  {Tarzia}(2021)}]{PhysRevB.103.104205}%
  \BibitemOpen
  \bibfield  {author} {\bibinfo {author} {\bibfnamefont {G.}~\bibnamefont
  {Biroli}}\ and\ \bibinfo {author} {\bibfnamefont {M.}~\bibnamefont
  {Tarzia}},\ }\bibfield  {title} {\bibinfo {title}
  {{L}\'evy-{R}osenzweig-{P}orter random matrix ensemble},\ }\href
  {https://doi.org/10.1103/PhysRevB.103.104205} {\bibfield  {journal} {\bibinfo
   {journal} {Phys. Rev. B}\ }\textbf {\bibinfo {volume} {103}},\ \bibinfo
  {pages} {104205} (\bibinfo {year} {2021})}\BibitemShut {NoStop}%
\bibitem [{\citenamefont {Kravtsov}\ \emph {et~al.}(2020)\citenamefont
  {Kravtsov}, \citenamefont {Khaymovich}, \citenamefont {Altshuler},\ and\
  \citenamefont {Ioffe}}]{kravtsov2020localization}%
  \BibitemOpen
  \bibfield  {author} {\bibinfo {author} {\bibfnamefont {V.~E.}\ \bibnamefont
  {Kravtsov}}, \bibinfo {author} {\bibfnamefont {I.~M.}\ \bibnamefont
  {Khaymovich}}, \bibinfo {author} {\bibfnamefont {B.~L.}\ \bibnamefont
  {Altshuler}},\ and\ \bibinfo {author} {\bibfnamefont {L.~B.}\ \bibnamefont
  {Ioffe}},\ }\href@noop {} {\bibinfo {title} {Localization transition on the
  random regular graph as an unstable tricritical point in a log-normal
  rosenzweig-porter random matrix ensemble}} (\bibinfo {year} {2020}),\ \Eprint
  {https://arxiv.org/abs/2002.02979} {arXiv:2002.02979 [cond-mat.dis-nn]}
  \BibitemShut {NoStop}%
\bibitem [{\citenamefont {Khaymovich}\ and\ \citenamefont
  {Kravtsov}(2021)}]{10.21468/SciPostPhys.11.2.045}%
  \BibitemOpen
  \bibfield  {author} {\bibinfo {author} {\bibfnamefont {I.~M.}\ \bibnamefont
  {Khaymovich}}\ and\ \bibinfo {author} {\bibfnamefont {V.~E.}\ \bibnamefont
  {Kravtsov}},\ }\bibfield  {title} {\bibinfo {title} {{{D}ynamical phases in a
  ``multifractal'' {R}osenzweig-{P}orter model}},\ }\href
  {https://doi.org/10.21468/SciPostPhys.11.2.045} {\bibfield  {journal}
  {\bibinfo  {journal} {SciPost Phys.}\ }\textbf {\bibinfo {volume} {11}},\
  \bibinfo {pages} {045} (\bibinfo {year} {2021})}\BibitemShut {NoStop}%
\bibitem [{\citenamefont {Garc\'{\i}a-Garc\'{\i}a}\ and\ \citenamefont
  {Wang}(2005)}]{PhysRevLett.94.244102}%
  \BibitemOpen
  \bibfield  {author} {\bibinfo {author} {\bibfnamefont {A.~M.}\ \bibnamefont
  {Garc\'{\i}a-Garc\'{\i}a}}\ and\ \bibinfo {author} {\bibfnamefont
  {J.}~\bibnamefont {Wang}},\ }\bibfield  {title} {\bibinfo {title} {{A}nderson
  {T}ransition in {Q}uantum {C}haos},\ }\href
  {https://doi.org/10.1103/PhysRevLett.94.244102} {\bibfield  {journal}
  {\bibinfo  {journal} {Phys. Rev. Lett.}\ }\textbf {\bibinfo {volume} {94}},\
  \bibinfo {pages} {244102} (\bibinfo {year} {2005})}\BibitemShut {NoStop}%
\bibitem [{\citenamefont {Chirikov}(1979)}]{CHIRIKOV1979263}%
  \BibitemOpen
  \bibfield  {author} {\bibinfo {author} {\bibfnamefont {B.~V.}\ \bibnamefont
  {Chirikov}},\ }\bibfield  {title} {\bibinfo {title} {{A} universal
  instability of many-dimensional oscillator systems},\ }\href
  {https://doi.org/https://doi.org/10.1016/0370-1573(79)90023-1} {\bibfield
  {journal} {\bibinfo  {journal} {Physics Reports}\ }\textbf {\bibinfo {volume}
  {52}},\ \bibinfo {pages} {263} (\bibinfo {year} {1979})}\BibitemShut
  {NoStop}%
\bibitem [{\citenamefont {Izrailev}(1990)}]{IZRAILEV1990299}%
  \BibitemOpen
  \bibfield  {author} {\bibinfo {author} {\bibfnamefont {F.~M.}\ \bibnamefont
  {Izrailev}},\ }\bibfield  {title} {\bibinfo {title} {{S}imple models of
  quantum chaos: {S}pectrum and eigenfunctions},\ }\href
  {https://doi.org/https://doi.org/10.1016/0370-1573(90)90067-C} {\bibfield
  {journal} {\bibinfo  {journal} {Physics Reports}\ }\textbf {\bibinfo {volume}
  {196}},\ \bibinfo {pages} {299} (\bibinfo {year} {1990})}\BibitemShut
  {NoStop}%
\bibitem [{\citenamefont {Chen}\ \emph {et~al.}(2024)\citenamefont {Chen},
  \citenamefont {Giraud}, \citenamefont {Gong},\ and\ \citenamefont
  {Lemarié}}]{LongPaper}%
  \BibitemOpen
  \bibfield  {author} {\bibinfo {author} {\bibfnamefont {W.}~\bibnamefont
  {Chen}}, \bibinfo {author} {\bibfnamefont {O.}~\bibnamefont {Giraud}},
  \bibinfo {author} {\bibfnamefont {J.}~\bibnamefont {Gong}},\ and\ \bibinfo
  {author} {\bibfnamefont {G.}~\bibnamefont {Lemarié}},\ }\bibfield  {title}
  {\bibinfo {title} {{In Preparation}},\ }\href@noop {} {\  (\bibinfo {year}
  {2024})}\BibitemShut {NoStop}%
\bibitem [{\citenamefont {Luitz}(2021)}]{luitz2021polynomial}%
  \BibitemOpen
  \bibfield  {author} {\bibinfo {author} {\bibfnamefont {D.~J.}\ \bibnamefont
  {Luitz}},\ }\bibfield  {title} {\bibinfo {title} {{Polynomial filter
  diagonalization of large Floquet unitary operators}},\ }\href
  {https://doi.org/10.21468/SciPostPhys.11.2.021} {\bibfield  {journal}
  {\bibinfo  {journal} {SciPost Phys.}\ }\textbf {\bibinfo {volume} {11}},\
  \bibinfo {pages} {021} (\bibinfo {year} {2021})}\BibitemShut {NoStop}%
\bibitem [{sup()}]{sup}%
  \BibitemOpen
  \href@noop {} {}\bibinfo {note} {See Supplemental Material.}\BibitemShut
  {Stop}%
\bibitem [{\citenamefont {Kravtsov}\ \emph {et~al.}(2012)\citenamefont
  {Kravtsov}, \citenamefont {Yevtushenko}, \citenamefont {Snajberk},\ and\
  \citenamefont {Cuevas}}]{PhysRevE.86.021136}%
  \BibitemOpen
  \bibfield  {author} {\bibinfo {author} {\bibfnamefont {V.~E.}\ \bibnamefont
  {Kravtsov}}, \bibinfo {author} {\bibfnamefont {O.~M.}\ \bibnamefont
  {Yevtushenko}}, \bibinfo {author} {\bibfnamefont {P.}~\bibnamefont
  {Snajberk}},\ and\ \bibinfo {author} {\bibfnamefont {E.}~\bibnamefont
  {Cuevas}},\ }\bibfield  {title} {\bibinfo {title} {{L}\'evy flights and
  multifractality in quantum critical diffusion and in classical random walks
  on fractals},\ }\href {https://doi.org/10.1103/PhysRevE.86.021136} {\bibfield
   {journal} {\bibinfo  {journal} {Phys. Rev. E}\ }\textbf {\bibinfo {volume}
  {86}},\ \bibinfo {pages} {021136} (\bibinfo {year} {2012})}\BibitemShut
  {NoStop}%
\bibitem [{\citenamefont {Ketzmerick}\ \emph {et~al.}(1997)\citenamefont
  {Ketzmerick}, \citenamefont {Kruse}, \citenamefont {Kraut},\ and\
  \citenamefont {Geisel}}]{PhysRevLett.79.1959}%
  \BibitemOpen
  \bibfield  {author} {\bibinfo {author} {\bibfnamefont {R.}~\bibnamefont
  {Ketzmerick}}, \bibinfo {author} {\bibfnamefont {K.}~\bibnamefont {Kruse}},
  \bibinfo {author} {\bibfnamefont {S.}~\bibnamefont {Kraut}},\ and\ \bibinfo
  {author} {\bibfnamefont {T.}~\bibnamefont {Geisel}},\ }\bibfield  {title}
  {\bibinfo {title} {{W}hat {D}etermines the {S}preading of a {W}ave
  {P}acket?},\ }\href {https://doi.org/10.1103/PhysRevLett.79.1959} {\bibfield
  {journal} {\bibinfo  {journal} {Phys. Rev. Lett.}\ }\textbf {\bibinfo
  {volume} {79}},\ \bibinfo {pages} {1959} (\bibinfo {year}
  {1997})}\BibitemShut {NoStop}%
\bibitem [{\citenamefont {Chalker}(1990)}]{CHALKER1990253}%
  \BibitemOpen
  \bibfield  {author} {\bibinfo {author} {\bibfnamefont {J.}~\bibnamefont
  {Chalker}},\ }\bibfield  {title} {\bibinfo {title} {{S}caling and
  eigenfunction correlations near a mobility edge},\ }\href
  {https://doi.org/https://doi.org/10.1016/0378-4371(90)90056-X} {\bibfield
  {journal} {\bibinfo  {journal} {Physica A: Statistical Mechanics and its
  Applications}\ }\textbf {\bibinfo {volume} {167}},\ \bibinfo {pages} {253}
  (\bibinfo {year} {1990})}\BibitemShut {NoStop}%
\end{thebibliography}%

\ifarXiv
    

\section*{Supplementary material}

# Supplemental Material for "Quantum logarithmic multifractality"

## I. NUMERICAL RESULTS FOR THE CORRELATION FUNCTION

In this section, we present numerical results for the correlation function

$$C(r) \equiv \langle \sum_{i=1}^{N} |\psi_j(i)|^2 |\psi_j(i+r)|^2 \rangle \qquad (S1)$$

of the SRBM model and the SRUM model. In (S1) we perform averaging on both disorder realizations and eigenstates $\psi_j$. The numerical results displayed in Fig. S1 and Fig. S2 support our argument of the spatial decay behavior

$$C(r) \sim \frac{1}{r(\ln r)^\alpha}. \qquad (S2)$$

We find that the decay exponent $\alpha$ is related to the log-multifractal dimension $d_2$ by $\alpha + d_2 \approx 1$ in the SRBM model. However, such a relationship is absent in the SRUM model.

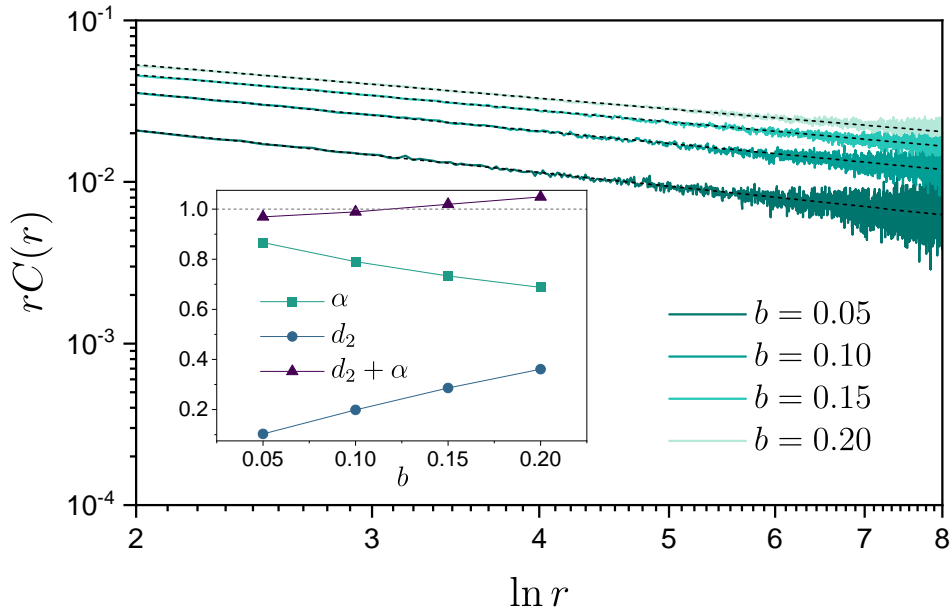

FIG. S1. Spatial decay of the correlation function $C(r)$ Eq. (S2) in the SRBM model for different $b$ values as indicated by the labels. The data is plotted as $rC(r)$ a function of $\ln r$ in log-log scale, the dash lines are power-law fits $rC(r) \sim (\ln r)^{-\alpha}$. Results have been averaged over $18,000$ realizations with $N = 2^{14}$. Inset: Verification of the relationship $d_2 + \alpha = 1$.

## II. NUMERICAL RESULTS FOR THE GENERALIZED SRBM MODEL

In this section, we present numerical results for the generalized SRBM model with parameter $\mu = 1/2$. From analytic considerations (see main text), we expect a localized behavior of the inverse participation ratio

$$\langle P_2 \rangle \sim (\ln N)^{-\mu} + \text{cst.} \qquad (S3)$$

and a spatial decay of the correlation function following

$$C(r) \sim \frac{1}{r(\ln r)^{1+\mu}}. \qquad (S4)$$

Numerical results presented in Fig. S3 validate the above predictions.

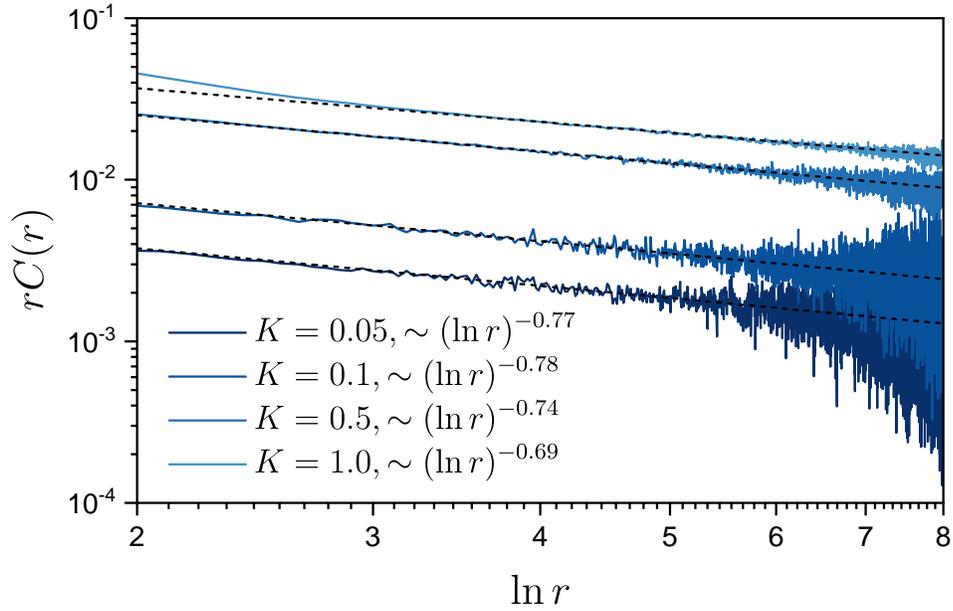

FIG. S2. Spatial decay for the correlation function $C(r)$ Eq. (S2) in the SRUM model for different $K$ values as indicated by the labels. The data is plotted as $rC(r)$ a function of $\ln r$ in log-log scale, the dash lines are power-law fittings $rC(r) \sim (\ln r)^{-\alpha}$. Results have been averaged over $36,000$ realizations with $N = 2^{16}$.

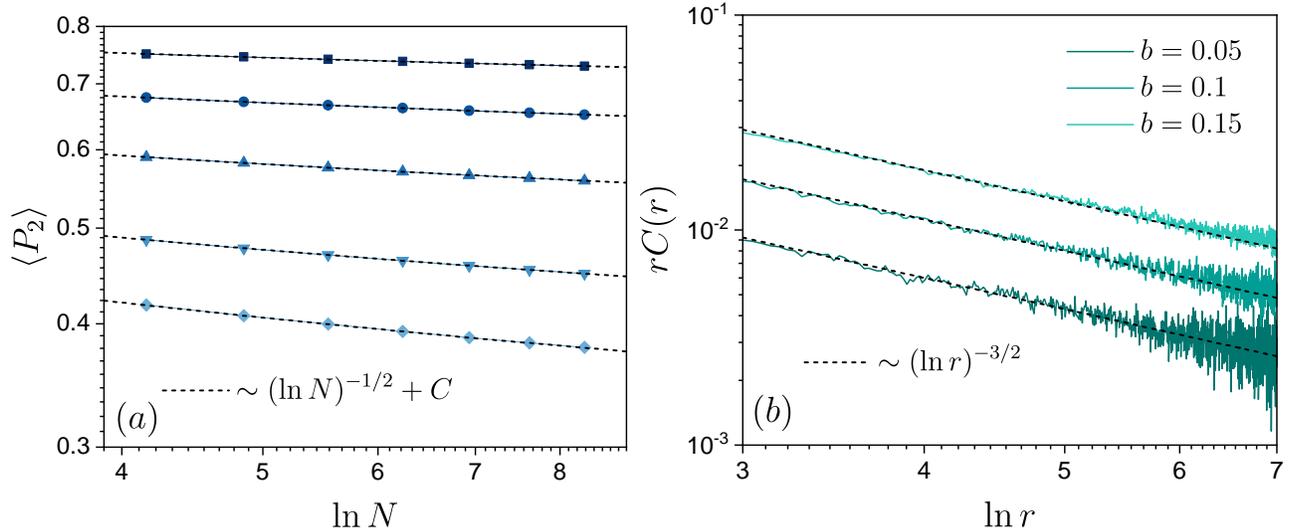

FIG. S3. (a) Scaling property Eq. (S3) for the generalized SRBM model with $\mu = 1/2$; different curves correspond to different $b$ values: $b = 0.05, 0.07, 0.1, 0.15, 0.2$ from top to bottom. The black dashed lines are power-law fits $\langle P_2 \rangle \sim (\ln N)^{-\mu} + C$. Disorder averaging ranges from $360,000$ realizations for $N = 2^6$ to $18,000$ realizations for $N = 2^{12}$. (b) Spatial decay for the correlation function $C(r)$ Eq. (S4) in the generalized SRBM model with $\mu = 1/2$ for different $b$ values as indicated by the labels. The data is plotted as $rC(r)$ a function of $\ln r$ in log-log scale, the dash lines are power-law fittings $rC(r) \sim (\ln r)^{-(1+\mu)}$. Results have been averaged over $18,000$ realizations with $N = 2^{14}$.


\fi

\end{document}